%% file: main.tex
\documentclass[letterpaper,twocolumn,10pt]{article}
\usepackage{deperated/usenix}

\usepackage{amsmath}

\usepackage{amssymb}
\usepackage[ruled,linesnumbered]{algorithm2e}
\usepackage{pifont}
\usepackage{enumitem}
\setlist[itemize]{leftmargin=*,topsep=0.2em,itemsep=0.2em,parsep=0pt,partopsep=0pt}
\usepackage{xspace}
\usepackage{array}
\usepackage{booktabs}
\usepackage{makecell}
\usepackage{multirow}
\usepackage[font=small,labelfont=bf]{caption}
\usepackage{float}
\usepackage{placeins}
\usepackage{soul}
\usepackage{tikz}
\usetikzlibrary{tikzmark}
\usepackage{eso-pic}

\definecolor{slotgray}{RGB}{224,224,224}
\sethlcolor{slotgray}
\soulregister{\texttt}{1}
\soulregister{\ref}{1}
\soulregister{\KwData}{1}

\newcommand{\slotbgleft}{-0.5em}
\newcommand{\slotbgright}{15em}
\newcommand{\slotbgtop}{2.2ex}
\newcommand{\slotbgbot}{-0.6ex}

\newcommand{\slotbgstart}[1]{%
  \tikzmark{slot#1start}%
  \expandafter\xdef\csname slotbg@l@#1\endcsname{\slotbgleft}%
  \expandafter\xdef\csname slotbg@r@#1\endcsname{\slotbgright}%
  \expandafter\xdef\csname slotbg@t@#1\endcsname{\slotbgtop}%
  \expandafter\xdef\csname slotbg@b@#1\endcsname{\slotbgbot}%
  \AddToShipoutPictureBG*{%
    \begin{tikzpicture}[overlay, remember picture]
      \fill[slotgray]
        ([xshift=\csname slotbg@l@#1\endcsname,
          yshift=\csname slotbg@t@#1\endcsname]pic cs:slot#1start) rectangle
        ([xshift=\csname slotbg@r@#1\endcsname,
          yshift=\csname slotbg@b@#1\endcsname]pic cs:slot#1end);
    \end{tikzpicture}%
  }%
}
\newcommand{\slotbgend}[1]{\tikzmark{slot#1end}}
\newcommand{\sysname}{LIOS\xspace}
\newcommand{\para}[1]{\noindent {\bf #1}}

\makeatletter
\def\@maketitle{\newpage
 \vbox to 1.25in{
 \vskip 0in
 \begin{center}%
  {\Large\bf \@title \par}%
  \vskip 0.10in minus 0.05in
  {\large\it
   \lineskip .5em
   \begin{tabular}[t]{c}\@author
   \end{tabular}\par}%
 \end{center}%
 \par
 \vspace*{\fill}
 }
}
\makeatother

\begin{document}
\sloppy
\date{}

\title{Leveraging I/O Stalls for Efficient Scheduling in ANNS}
\author{
{\rm Juncheng Zhang$^{1}$, Yuanming Ren$^{1}$, Yongkun Li$^{2}$, and Patrick P.C. Lee$^{1}$}\\
$^{1}$The Chinese University of Hong Kong \quad
$^{2}$University of Science and Technology of China
}

\maketitle

\begin{abstract}
Disk-based graph indexes for approximate nearest neighbor search (ANNS) must serve latency-sensitive queries and throughput-demanding updates concurrently. We observe that over 40\% of search-thread CPU time is spent stalling on disk I/O; such idle cycles are invisible to thread-level scheduling yet available for other work. We present \sysname (\underline{L}everage \underline{I}/\underline{O} \underline{S}tall), a framework that executes index updates inside search-side I/O stall windows. \sysname introduces three techniques: (i) splitting each update into resumable subtasks small enough to fit within a single stall window; (ii) bounding the expected overrun of update subtasks to a given threshold; and (iii) dynamically adjusting the fraction of idle time devoted to updates to drive end-to-end search latency degradation toward a user-specified target. We integrate \sysname into two update-optimized ANNS systems, FreshDiskANN and OdinANN. \sysname achieves speedups of up to 2.68$\times$ in insertion and 2.18$\times$ in deletion, with search latency degradation maintained near the user-specified target.
\end{abstract}

\begin{sloppypar}
\input{1-introduction}
\input{2-background}
\input{3-motivation}
\input{4-design}
\input{5-implementation}

\input{6-evaluation}
\input{7-relatedwork}
\input{8-conclusion}
\end{sloppypar}

\bibliographystyle{plain}
\bibliography{ref}

\clearpage
\appendix
\input{appendix}

\end{document}

%% file: 1-introduction.tex
\section{Introduction}
\label{sec:intro}

On-disk graph-based indexes \cite{malkov18,fu19,jayaram19,singh21,guo25,guo26} are widely used for large-scale Approximate Nearest Neighbor Search (ANNS) \cite{wang21,guo20,johnson19,sun23} in recommendation, web search, and retrieval-augmented generation \cite{lewis20}. As datasets grow to billions of vectors and continue to be ingested, these systems must concurrently handle two fundamentally different workloads: {\em latency-sensitive search queries} and {\em throughput-demanding index updates} \cite{singh21,guo26,xu23}. At the thread level, balancing the two under a shared CPU budget is difficult: devoting more cores to one workload reduces the CPU available to the other. Insufficient resources for updates cause the ANNS index to lag behind incoming data, degrading retrieval accuracy and result freshness, while insufficient resources for search directly degrade user-facing latency. Existing systems treat the two as competing consumers of a fixed resource pool, with no principled mechanism to reduce this tension. 

We observe that even when every search thread is continuously saturated with queries, over 40\% of its CPU time is stalled on I/O (\S\ref{sec:moti}). This \emph{intra-request} idle time is invisible to thread-level scheduling: the thread is actively processing a query, yet the CPU is idle until the disk read completes. If these idle cycles can be harvested and redirected to update tasks, it becomes possible to improve update throughput without reducing the CPU budget available to search, at a granularity that thread-level approaches cannot reach.

Applying this idea in practice faces several challenges. First, a single update operation typically lasts much longer than the idle intervals available during search I/O stalls, making fine-grained task decomposition essential to fit update work within short idle periods without interfering with search. Second, it is non-trivial to determine when and how much update work can be safely scheduled, as I/O stall durations are inherently variable and further distorted by techniques such as dynamic I/O width and user-space caching \cite{guo25,wang24}. Third, even when update tasks complete within the estimated idle window, search-update co-execution degrades search latency due to workload- and runtime-dependent effects (e.g., cache pollution and varying workload sensitivity) that cannot be modeled analytically in advance. Keeping degradation within the user's tolerance requires a runtime mechanism to monitor and adapt to these effects.

We present \sysname (\underline{L}everage \underline{I}/\underline{O} \underline{S}tall), a framework that co-executes search and update operations in graph-based ANNS systems by exploiting I/O idle cycles. \sysname introduces three key techniques: (i) \emph{fine-grained update decomposition}, which breaks monolithic update operations into resumable, checkpoint-able subtasks at both inter-vector and intra-vector granularities so that each piece fits within a short I/O stall window; (ii) \emph{overrun-bounded time budgeting}, which estimates from recent idle-time samples how long each update subtask can safely be executed within an idle window, bounding the expected overrun of update subtasks to a controlled fraction of the mean idle time; and (iii) \emph{adaptive utilization tuning}, which uses a feedback loop to dynamically tune the fraction of the idle-time budget devoted to updates, driving end-to-end search latency degradation toward a user-specified target $\theta$ (which is the only user-configurable parameter to maintain operational simplicity). \sysname is optimized with system-level mechanisms (\S\ref{sec:imple}) to achieve high search-update co-execution performance.

We integrate \sysname into two state-of-the-art update-optimized ANNS systems, FreshDiskANN \cite{singh21} and OdinANN \cite{guo26}. We evaluate \sysname on real-world datasets ranging from 10M to 500M vectors across two servers with distinct hardware configurations. At the 100M scale, \sysname achieves update speedups of up to 1.46$\times$ (insert) and 1.48$\times$ (delete) under FreshDiskANN, and up to 1.72$\times$ (insert) and 1.49$\times$ (delete) under OdinANN. At the 10M scale, the gains are even larger, reaching up to 2.68$\times$ (insert) and 2.18$\times$ (delete). Meanwhile, \sysname maintains controlled degradation in search latency. We will open-source our \sysname prototype in the final version of the paper.

%% file: 2-background.tex
\section{Background}
\label{sec:bg}

\subsection{Basics of Graph-based ANNS Systems}
\label{subsec:basic-of-graph-ann}

\begin{figure}[!t]
\centering
\includegraphics[width=0.98\columnwidth]{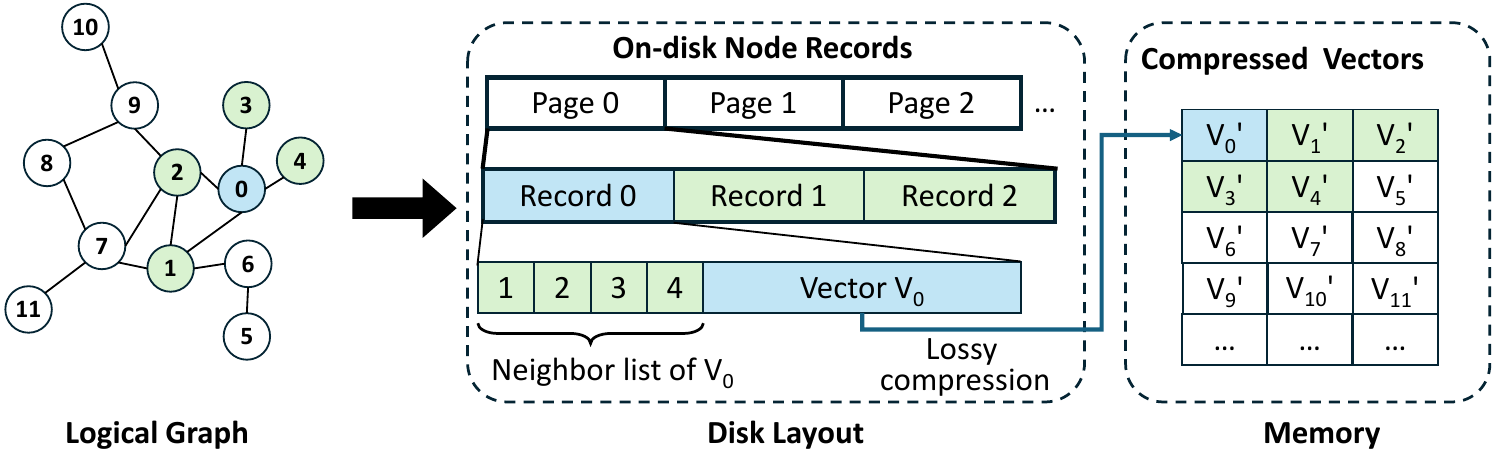}
\caption{DiskANN organizes its index by storing raw vectors together with their out-neighbors on disk, while keeping a compressed representation of the vectors in memory to enable fast similarity comparisons.}
\label{fig:disk-index-overview}
\end{figure}

We consider graph-based ANNS systems for vector storage in disk-based scenarios where datasets exceed memory capacity. Vectors are organized as {\em nodes} in a directed neighbor graph, enabling efficient similarity search through iterative graph traversal. DiskANN \cite{jayaram19} stands out as a pioneering graph-based ANNS system that has inspired subsequent optimizations, including FreshDiskANN \cite{singh21}, PipeANN \cite{guo25}, and OdinANN \cite{guo26}. Thus, we use DiskANN as the representative disk-based graph-based ANNS system for our discussion.

As shown in Figure~\ref{fig:disk-index-overview}, DiskANN organizes vectors as a \emph{logical graph}, in which each vector is a node and each directed edge points to an out-neighbor. On disk, this logical graph is materialized as fixed-size \emph{node records}: each record corresponds to a single vector and contains its raw data along with the IDs of its out-neighbors. The number of out-neighbors per node is bounded by a fixed upper limit $R$, so all records share a consistent size for alignment and page management. In memory, DiskANN keeps compressed vectors, typically generated using Product Quantization \cite{jegou10} or RaBitQ \cite{gao24}, to support fast approximate distance estimation.

\subsection{Search Algorithm}
\label{subsec:search-algo}


DiskANN's search procedure, commonly referred to as \emph{BeamSearch}, follows an iterative greedy strategy on the neighbor graph. Given a query vector $q$, the algorithm begins from a designated entry point and maintains two data structures: a candidate pool of maximum size $L$ tracking the most promising vectors seen so far, and a visited set recording vectors already expanded. Each iteration proceeds in two steps: \ding{182} it selects the $W$ closest unvisited candidates from the pool (where $W$ is the beam width), reads their on-disk records, and retrieves the out-neighbor lists; \ding{183} it computes distances between $q$ and all newly discovered neighbors using the compressed in-memory vectors, and inserts those neighbors into the candidate pool. If the pool exceeds size $L$, it is truncated to retain only the $L$ nearest candidates. This process repeats until every vector in the candidate pool has been visited, at which point the top-$K$ nearest vectors are returned.

PipeANN \cite{guo25} relaxes the strict per-hop data dependency in graph traversal by overlapping disk I/O with computation via pipelining; we refer to its search procedure as \emph{PipeSearch}, in contrast to the standard BeamSearch. Despite this optimization, the overall performance remains bounded by disk access latency. Each graph-traversal hop requires at least one disk read, and the per-hop disk I/O latency far exceeds the corresponding computation time, making disk access the dominant factor even with pipelining. In \S\ref{sec:moti}, we provide a quantitative breakdown of this I/O-bound characteristic.

\subsection{Update Algorithm}
\label{subsec:update_algorithm}


Supporting updates, including insertions and deletions, is crucial for maintaining the freshness of disk-based ANNS systems. FreshDiskANN \cite{singh21} buffers insertions in an auxiliary in-memory index while tracking pending deletions in a separate buffer. Once the in-memory insertion index grows beyond a threshold, a \emph{StreamingMerge} procedure is triggered to flush all buffered updates into the on-disk index. OdinANN \cite{guo26} adopts a similar buffering strategy for deletions, but applies {\em in-place} insertions directly into the on-disk graph, eliminating the in-memory index to reduce memory overhead and mitigate interference with search.

Despite the different approaches, both systems share a computationally intensive step: the neighbor selection procedure, commonly referred to as \emph{Prune}. After each update, Prune recomputes the neighbor sets of affected nodes to maintain the Sparse Neighborhood Graph (SNG) property \cite{arya1993}, which ensures the selected neighbors provide good angular coverage around each node while avoiding redundant short-range edges. Specifically, given a target point $p$, a candidate set $V$ of potential neighbors, a distance threshold $\alpha$, and a degree bound $R$, Prune iteratively selects up to $R$ neighbors. In each iteration, it picks the candidate $p^{*}$ closest to $p$ from $V$ and adds $p^{*}$ to the neighbor set. It then scans the remaining candidates and removes every $p'$ for which $\alpha \cdot d(p^{*}, p') \le d(p, p')$, filtering out candidates whose direction from $p$ is already covered by $p^{*}$. This process repeats until either $R$ neighbors have been selected or the candidate set is exhausted, yielding a neighbor set with diverse angular coverage around $p$ that respects the degree bound.

OdinANN proposes an improved variant, \emph{DeltaPrune}, to reduce pruning overhead, yet pruning remains the key computational bottleneck for updates, as we analyze in \S\ref{sec:moti}.


%% file: 3-motivation.tex
\section{Motivation}
\label{sec:moti}

We analyze two complementary phenomena: the CPU idle time that accumulates during search I/O stalls (\S\ref{subsec:search-idle}), and the high computational cost that dominates index updates (\S\ref{subsec:update-comp}).

\subsection{Unexploited Idle Time in Search}
\label{subsec:search-idle}

\begin{figure}[!t]
\centering
\setlength{\tabcolsep}{2pt}
\renewcommand{\arraystretch}{1.0}
\begin{tabular}{@{\ }cc}
\multicolumn{2}{c}{\includegraphics[width=3.25in]{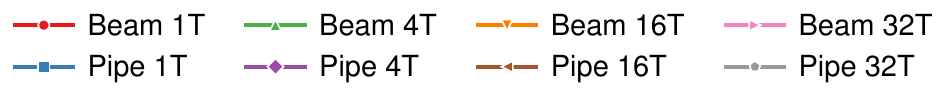}} \\

\includegraphics[width=1.62in]{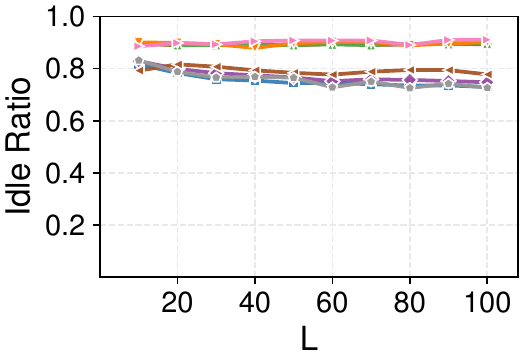} &
\includegraphics[width=1.499in]{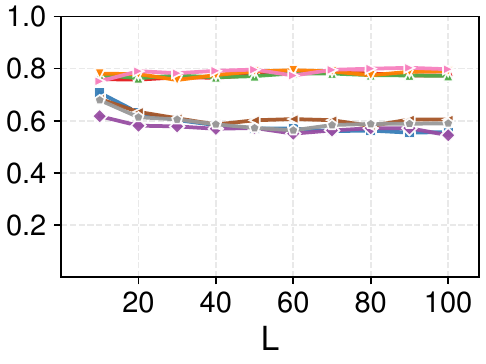} 
\vspace{-3pt}\\
\parbox[c]{1.62in}{\centering\small (a) SIFT-10M, $R=32$} &
\parbox[c]{1.499in}{\centering\small (b) SIFT-10M, $R=100$} 
\vspace{3pt}\\
\includegraphics[width=1.62in]{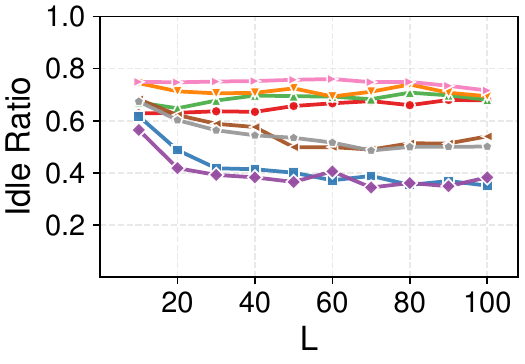} &
\includegraphics[width=1.499in]{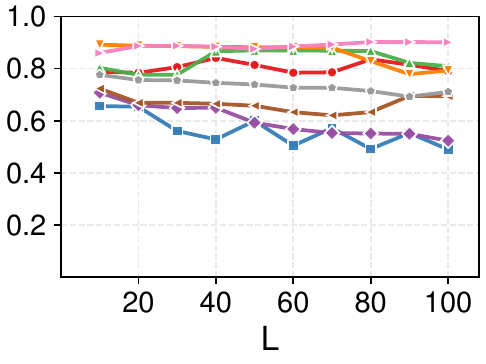} 
\vspace{-3pt}\\
\parbox[c]{1.62in}{\centering\small (c) SIFT-100M, $R=100$} &
\parbox[c]{1.499in}{\centering\small (d) SIFT-1B $R=64$}
\end{tabular}
\vspace{-9pt}
\caption{Idle ratios of BeamSearch and PipeSearch across dataset scales and index configurations.} 
\label{fig:diskann-idle-time}
\vspace{-6pt}
\end{figure}

Search threads in graph-based ANNS systems spend a substantial fraction of their time waiting on disk I/O, consistently across different graph indexes and system configurations. We define the {\em idle ratio} as the fraction of a search thread's execution time spent waiting for outstanding disk reads to complete (i.e., the time during which the thread holds the CPU but performs no useful computation). We validate the prevalence of this idle time by profiling BeamSearch and PipeSearch across multiple datasets and workloads. We consider datasets at scales of 10M, 100M, and 1B vectors. We fix the beam width $W=4$ (\S\ref{subsec:basic-of-graph-ann}), configure different $R$ (i.e., the maximum out-degree per node;  \S\ref{subsec:basic-of-graph-ann}) and $L$ (i.e., the maximum number of candidates; \S\ref{subsec:search-algo}), and vary the thread count from 1 to 32. 

Figure~\ref{fig:diskann-idle-time} shows the idle ratios across different settings.  BeamSearch exhibits a high idle ratio of 60-80\%. This is because the set of candidate nodes to fetch in hop $h+1$ is determined by the distance computations on hop $h$'s results, so each hop creates a strict sequential data dependency that forces the CPU to wait for the current disk read before it can issue the next one. PipeSearch achieves lower idle ratios by pipelining computation with disk I/O, allowing distance calculations to proceed while the next batch of data is being fetched. However, PipeSearch's idle ratio remains above 40\%, since the per-hop I/O latency still significantly exceeds the computation time (a condition inherent to disk-based traversal); even with the next hop's I/O already in flight, the CPU completes its distance computations well before the data arrives, leaving it idle for the residual gap. Thus, a pipelined search thread always carries residual idle intervals.

Note that simple thread-level optimizations, such as adding search threads or reallocating cores, cannot readily reclaim this idle time. A stalled search thread mid-query occupies its CPU context and cannot be preempted with another query without incurring context-switch overhead and breaking query-level state. The stall is {\em intra-request} (i.e., occurring within an actively executing query) and is hence invisible to thread-level or queue-based scheduling policies.

\para{Observation \#1.} {\em Search threads in graph-based ANNS systems spend over 40\% of time stalled on per-hop disk I/O, even with pipelined execution. This idle time is intra-request and cannot be readily recovered by thread-level scheduling.}

\subsection{High Computation Cost of Updates}
\label{subsec:update-comp}

\begin{figure}[!t]
\centering
\setlength{\tabcolsep}{2pt}
\renewcommand{\arraystretch}{1.0}
\begin{tabular}{@{\ }cc}
\multicolumn{2}{c}{\includegraphics[width=2.15in]{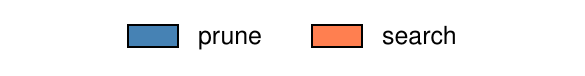}} \\[-6pt]

\includegraphics[width=1.80in]{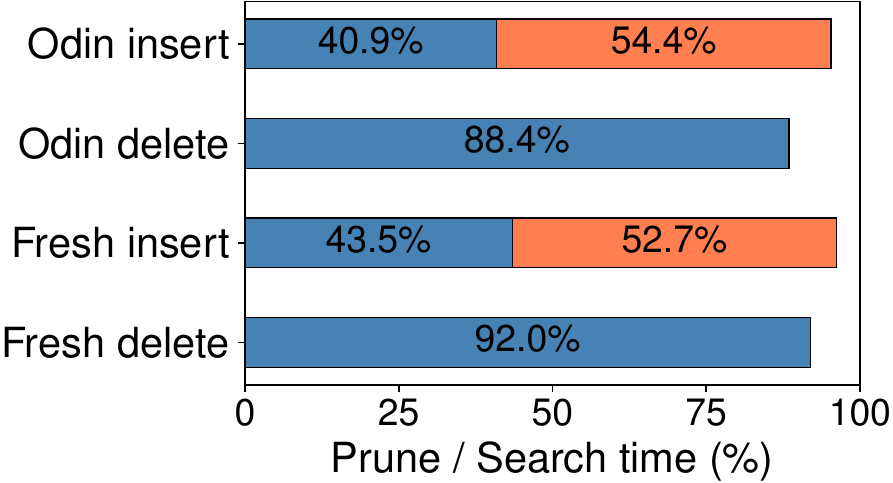} &
\includegraphics[width=1.32in]{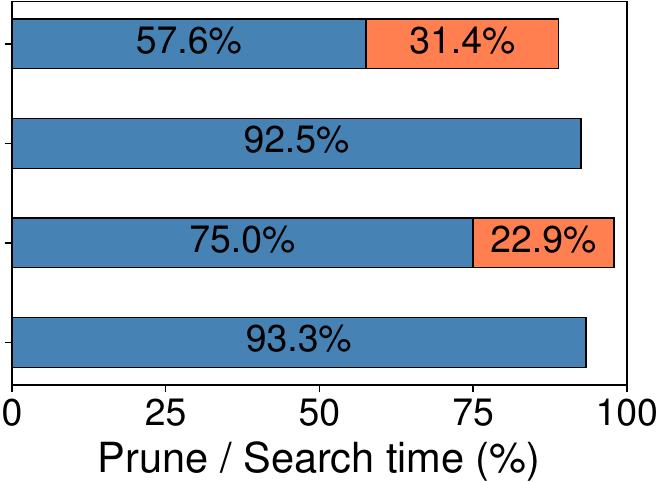} \\
\parbox[c]{1.80in}{\centering\small (a) $(R,L)=(32,32)$} &
\parbox[c]{1.32in}{\centering\small (b) $(R,L)=(100,100)$} 
\end{tabular}
\vspace{-9pt}
\caption{Proportion of prune and search times in the overall update time in SIFT-10M.}
\label{fig:prune-overhead}
\vspace{-6pt}
\end{figure}

We profile the insert and delete operations of FreshDiskANN \cite{singh21} and OdinANN \cite{guo26}, the two state-of-the-art update-optimized graph-based ANNS systems (\S\ref{subsec:update_algorithm}). Figure~\ref{fig:prune-overhead} shows their update time breakdown on the SIFT-10M dataset under $(R,L)=(32,32)$ and $(R,L)=(100,100)$. 

For insertions, the total cost is dominated by two components: (i) the search phase, which locates candidate neighbors for the new vector, and (ii) the subsequent prune phase, which selects a neighbor subset satisfying the SNG property (\S\ref{subsec:update_algorithm}). Under $(R, L) = (32, 32)$, pruning accounts for 40.9\% of insertion time in OdinANN and 43.5\% in FreshDiskANN. Increasing the graph degree to $(R, L) = (100, 100)$ raises these ratios to 57.6\% and 75.0\%, respectively, as a larger $R$ incurs more iterations and distance computations during neighbor selection. Although OdinANN's DeltaPrune reduces the absolute pruning cost of FreshDiskANN, pruning remains a major bottleneck, consuming 40.9–57.6\% of insertion time.

For deletions, the computational burden is even higher. Unlike insertions, which require pruning only the newly inserted node and its displaced neighbors, deletions must identify every node in the graph whose neighbor list references a deleted vector, and each such node must undergo local pruning to repair its neighbor set. This graph-wide repair causes the prune ratio for deletions to far exceed that for insertions. Across both $R=32$ and $R=100$ configurations on SIFT-10M, pruning accounts for 88.4–93.3\% of the total deletion time for OdinANN and FreshDiskANN, confirming that update throughput is constrained by CPU computations rather than I/O.

\para{Observation \#2.} {\em Index updates in graph-based ANNS systems are bottlenecked by CPU-intensive neighbor pruning, which accounts for 40.9–93.3\% of total update time across both insertions and deletions.}

%% file: 4-design.tex
\section{\sysname Design}
\label{sec:design}

\begin{figure}[!t]
\centering
\includegraphics[width=0.97\linewidth]{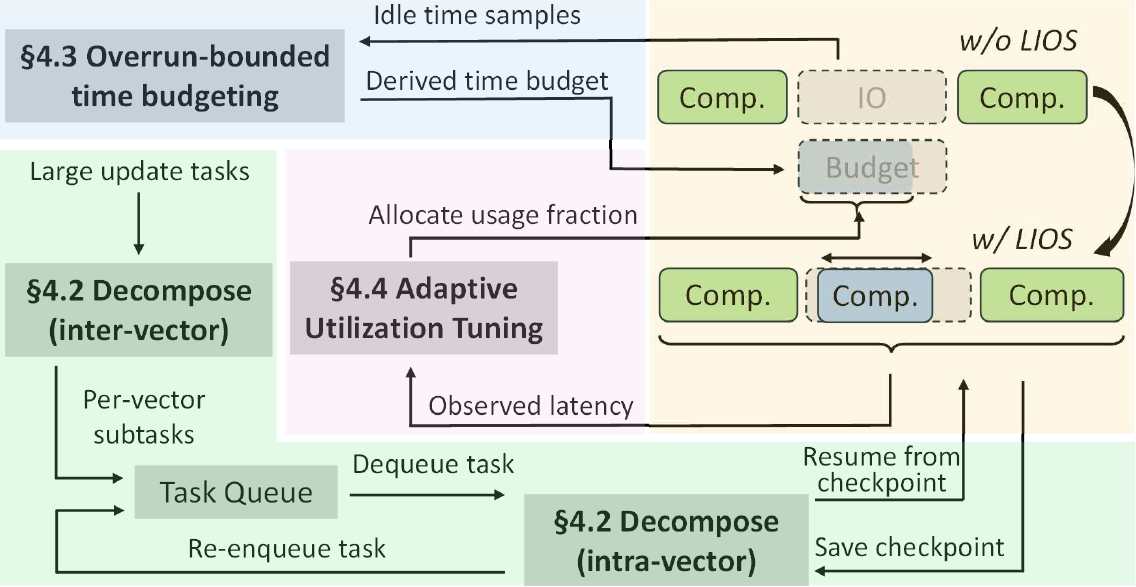}
\caption{Overview of \sysname's co-execution design.}
\label{fig:design-overview}
\end{figure}

\subsection{Overview}
\label{subsec:design-overview}

Our observations in \S\ref{sec:moti} reveal a natural complementarity: search threads spend over 40\% of their CPU time stalling on disk I/O, while updates are bottlenecked by CPU-intensive neighbor pruning. This motivates us to harvest idle CPU cycles within search threads and redirect them to update computation, improving update throughput without reducing the CPU budget available to search. However, realizing this idea in practice introduces three challenges:
\begin{itemize}[leftmargin=*,topsep=0.2em,itemsep=0.2em,parsep=0pt,partopsep=0pt]
\item
{\em Granularity mismatch}: A single update operation (e.g., pruning a vector's neighbor set) typically lasts much longer than the idle intervals available in a single search hop, so naively executing updates during I/O stalls would spill over into the search computation phase and inflate query latency. 
\item
{\em Variable and unpredictable idle periods:} Idle intervals vary unpredictably across hops, and optimizations such as dynamic I/O width and user-space caching further distort the underlying distribution, making it difficult to determine how much update work can be safely scheduled in any given window. 
\item
{\em Bounding end-to-end search latency degradation:} Co-execution degrades end-to-end search latency through workload- and runtime-dependent effects (e.g., cache pollution from displaced search data and varying sensitivity across workload phases) that cannot be modeled analytically in advance. Keeping the degradation within the user's tolerance requires a runtime mechanism that observes and adapts to these effects.
\end{itemize}

\sysname is a search-update co-execution framework designed to address these challenges in graph-based ANNS systems.  It introduces three techniques: 
\begin{itemize}[leftmargin=*,topsep=0.2em,itemsep=0.2em,parsep=0pt,partopsep=0pt]
\item
{\em Fine-grained update decomposition} (\S\ref{subsec:update-decomposition}), which breaks monolithic updates into independent per-vector subtasks, and replaces the original pruning procedure within each subtask with a resumable, checkpoint-able procedure that can pause and resume across multiple idle windows; 
\item
{\em Overrun-bounded time budgeting} (\S\ref{subsec:idle-time-estimation}), which derives a per-hop time budget from empirical idle time data that bounds the expected overrun (i.e., the time by which a scheduled update task exceeds the actual idle window) to a controlled fraction of the mean idle time by leveraging the temporal stability of short-term idle time distributions;
\item
{\em Adaptive utilization tuning} (\S\ref{subsec:utuner}), which dynamically adjusts the fraction of the time budget allocated to updates via a feedback-driven tuner, thereby steering end-to-end search latency degradation within a user-specified target.
\end{itemize}

\sysname exposes only one user-facing parameter, $\theta$, which specifies the allowed ratio of end-to-end search latency degradation. It guarantees search performance by bounding I/O-window overruns (\S\ref{subsec:idle-time-estimation}) and controlling co-execution overhead (\S\ref{subsec:utuner}). Both sources of degradation are regulated by $\theta$ to provide end-to-end latency guarantees.

Figure~\ref{fig:design-overview} depicts \sysname's workflow. \sysname first decomposes a large update task into independent per-vector subtasks, which are then enqueued into a FIFO task queue. During each search hop, \sysname derives an overrun-bounded budget from recent idle-time samples and allocates an online-tuned fraction of the derived budget to updates. It dequeues a subtask and executes the subtask within the resulting time slice. If the budget expires before the subtask completes (e.g., mid-pruning), the resumable pruning mechanism checkpoints the task's progress and re-enqueues the remaining work for completion in a subsequent idle window.

\subsection{Decomposition of Update Tasks}
\label{subsec:update-decomposition}

\begin{figure}[!t]
\centering
\begin{tabular}{@{\ }c@{\ }c}
\includegraphics[width=1.37in]{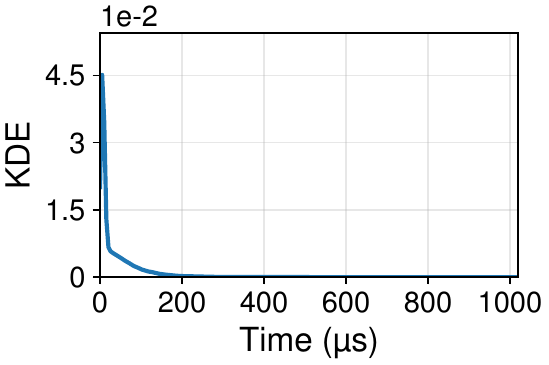} &
\includegraphics[width=1.36in]{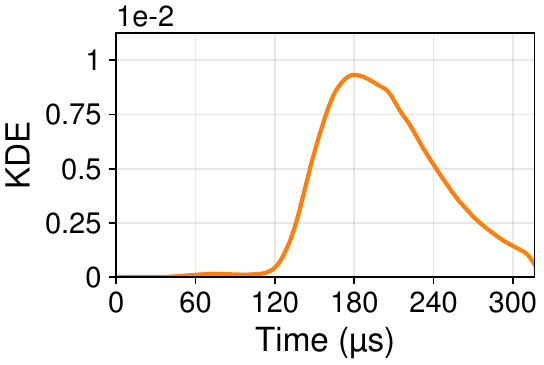}
\\
{\small (a) PipeSearch idle} &
{\small (b) BeamSearch idle}
\\
\includegraphics[width=1.37in]{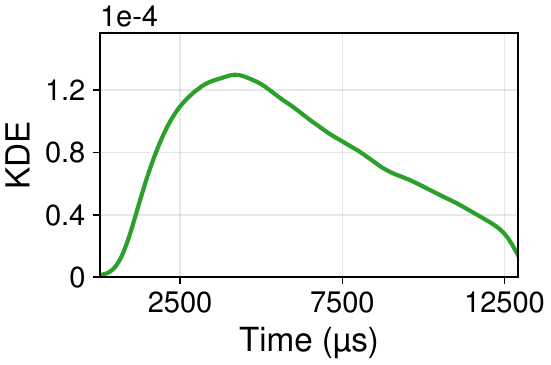} &
\includegraphics[width=1.34in]{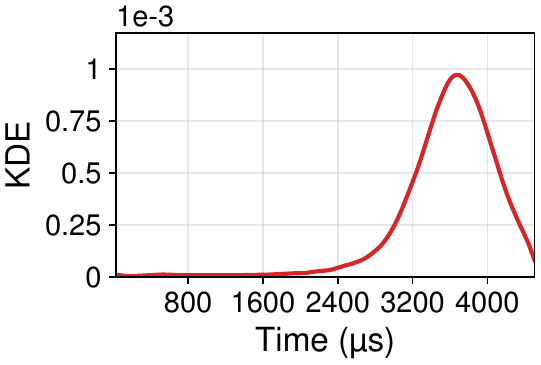}
\\
{\small (c) Delete per-node} & 
{\small (d) Insert per-node}
\end{tabular}
\vspace{-9pt}
\caption{KDE plots of per-hop search idle (a--b) and per-vector update computation time (c--d) on SIFT-10M, $R{=}100$.}
\label{fig:mismatch-idle-computation-time}
\vspace{-14pt}
\end{figure}

\sysname decomposes update work at two levels: {\em inter-vector decomposition} partitions an update task into independent per-vector subtasks, while {\em intra-vector decomposition} enables pruning inside each task resumable across idle windows.

\para{Inter-vector decomposition.} An update operation in graph-based ANNS can be decomposed into a collection of per-vector neighbor-set updates: each affected vector must have its neighbor list recomputed via the Prune procedure (\S\ref{subsec:update_algorithm}). For insertions, the affected set consists of the inserted vector itself and the $O(R)$ vectors in its reverse neighborhood whose neighbor lists may now include the new vector as a candidate. For deletions, identifying the affected set requires first scanning the graph to find all vectors whose neighbor lists reference a deleted entry; once this identification phase is complete and tasks are enqueued, each affected vector's pruning computation depends only on its own local candidate set and distances, and is hence independent of all other per-vector tasks. This independence means that per-vector tasks can be executed in any order without affecting the final graph, and it allows \sysname to partition a high-level update into a set of per-vector subtasks, which are enqueued into a shared task queue (\S\ref{sec:imple}).

\para{Intra-vector decomposition.} Inter-vector decomposition alone is insufficient, since a single vector's pruning time can far exceed an idle window. Figure~\ref{fig:mismatch-idle-computation-time} shows the idle-time and update-time distributions, using kernel density estimation (KDE) \cite{parzen1962}, a non-parametric method for estimating probability densities. The per-hop idle time is typically below 300\,\textmu s (Figures~\ref{fig:mismatch-idle-computation-time}(a)-\ref{fig:mismatch-idle-computation-time}(b)), whereas per-vector updates take roughly 800--4,500\,\textmu s for insertions and 1,000--12,500\,\textmu s for deletions (Figures~\ref{fig:mismatch-idle-computation-time}(c)-\ref{fig:mismatch-idle-computation-time}(d)), often exceeding idle windows by an order of magnitude. Within each per-vector update, the dominant cost is the Prune procedure, which accounts for 40--93\% of total update time (\S\ref{subsec:update-comp}), while the remaining computations (candidate preparation, distance computation, and adjacency-list finalization) are lightweight enough to complete within a single idle window and need not be split further. Thus, \sysname proposes \emph{resumable pruning}, which transforms the original Prune into a checkpoint-able procedure, enabling it to pause mid-execution and resume in a subsequent idle window.

\begin{algorithm}[t]
\SetArgSty{textnormal}
\newcommand{\algid}[1]{\mbox{\texttt{#1}}}
\newcommand{\algarr}[2]{\algid{#1}[#2]}
\caption{Resumable Pruning}
\label{alg:resumable-prune}
\tcc*[h]{Shaded code represents \sysname's logic added to the original Prune}\\
\KwIn{Point $p$, sorted candidate list \texttt{pool}, distance threshold $\alpha$, degree bound $R$, \hl{time budget $\tau_{\text{est}}$}}
\KwData{\hl{Checkpoint: \texttt{result}, \texttt{done}$[0..\text{len}-1]$, $i$, $j$}}
\KwOut{Pruned neighbor set \texttt{result}}
{%
\renewcommand{\slotbgleft}{-1.0em}%
\renewcommand{\slotbgright}{16.6em}%
\renewcommand{\slotbgtop}{1.8ex}%
\renewcommand{\slotbgbot}{-0.8ex}%
\slotbgstart{resume}\nlset{1}\textbf{if} resuming from checkpoint \textbf{then}\;
\Indp
restore \algid{result}, \algid{done}, $i$, $j$\;
\textbf{goto} line~\ref{line:inner}\slotbgend{resume}\;
\Indm
}
$\algid{result} \leftarrow \emptyset$; $\algid{done}[0..\algid{len}-1] \leftarrow \algid{false}$; $i \leftarrow 0$\;
\For{$i \leftarrow 0$ \KwTo $|\algid{pool}|-1$}{
    \If{$\algarr{done}{i} = \algid{true}$}{\textbf{continue}\;}
    $\algid{result} \leftarrow \algid{result} \cup \{\algarr{pool}{i}\}$; $\algarr{done}{i} \leftarrow \algid{true}$\;
    \lIf{$|\algid{result}| \ge R$}{\Return \algid{result}}
    \For(\label{line:inner}){$j \leftarrow i+1$ \KwTo $|\algid{pool}|-1$}{
        \If{$\algarr{done}{j} = \algid{false}$ \textbf{and} $\alpha \cdot d(\algarr{pool}{i},\, \algarr{pool}{j}) \le d(p,\, \algarr{pool}{j})$}{
            $\algarr{done}{j} \leftarrow \algid{true}$\;
        }
        {%
        \renewcommand{\slotbgleft}{-4.3em}%
        \renewcommand{\slotbgright}{9.5em}%
        \renewcommand{\slotbgtop}{1.8ex}%
        \renewcommand{\slotbgbot}{-0.8ex}%
        \slotbgstart{budget}\nlset{15}\textbf{if} elapsed time $\ge \tau_{\text{est}}$ \textbf{then}\;
        \Indp
        save checkpoint (\algid{result}, \algid{done}, $i$, $j{+}1$)\;
        \textbf{yield} to search thread\slotbgend{budget}\;
        \Indm
        }
    }
}
\Return \algid{result}\;
\end{algorithm}

\para{Resumable pruning.} Algorithm~\ref{alg:resumable-prune} shows the resumable pruning procedure, which builds upon the original Prune procedure (the shaded text marks \sysname's additions). \sysname recasts Prune as a resumable procedure whose checkpoint state comprises four mutable variables: the current result set \texttt{result}, a per-candidate flag \texttt{done} (indicating whether a candidate has been selected or eliminated under the SNG criterion), and two loop cursors $i$ and $j$. The candidate list \texttt{pool} is immutable and fixed at task creation time, so it does not need to be checkpointed.
When the elapsed time reaches the time budget $\tau_{\text{est}}$ derived from overrun-bounded budgeting (\S\ref{subsec:idle-time-estimation}), the procedure saves the checkpoint and yields control to the search thread. Upon resumption, pruning restores the checkpoint and continues from the saved $j$, producing a result that is bit-identical to that of uninterrupted execution.

The correctness of resumable pruning follows from two properties. First, since the checkpoint captures all four mutable variables, resumed execution is fully deterministic. Second, each per-vector task has its own \texttt{pool}, which is never modified by other threads, so no synchronization is required.

\para{Example.} Consider pruning a target point with degree bound $R=3$ and a candidate pool of 6 vectors sorted by distance: \texttt{pool} $= [c_0, c_1, \ldots, c_5]$, with \texttt{done} $= [F, F, F, F, F, F]$ ($F$ denotes false and $T$ denotes true), \texttt{result} $= \emptyset$, and both loop cursors $i$ and $j$ start at 0.

In the first idle window, the outer loop finds $c_0$ (non-done at $i=0$), selects it as a neighbor, adds $c_0$ to \texttt{result}, and marks \texttt{done}[0] $= T$. The inner loop then evaluates $c_1$ through $c_5$ against $c_0$: $c_1$ and $c_3$ fail the SNG criterion, so \texttt{done}[1] and \texttt{done}[3] are also set to $T$. After the inner loop completes, the outer loop advances to $i=2$ (skipping $i=1$ since \texttt{done}[1] $= T$), selects $c_2$, adds it to \texttt{result}, and marks \texttt{done}[2] $ = T$. The inner loop now evaluates $c_3$, $c_4$, and $c_5$ against $c_2$: at $j=4$, the SNG check does not eliminate $c_4$, and the subsequent budget check fires, signaling that the time budget has expired. The checkpoint (i.e., \texttt{result}=$\{c_0, c_2\}$, \texttt{done} $= [T,T,T,T,F,F]$, $i=2$, $j=5$) is saved, and control returns to the search thread.

In the next idle window, pruning restores the checkpoint and resumes the inner loop from $j=5$: it checks $c_5$ against $c_2$, which is not eliminated. The inner loop completes, and the outer loop advances to find the next non-done candidate: $i=3$ is skipped (\texttt{done}[3] $=T$), and $c_4$ is selected at $i=4$ as the third neighbor and added to \texttt{result}. Now, we have |\texttt{result}| $= R = 3$, and the pruning terminates. The final neighbor set $\{c_0, c_2, c_4\}$ is identical to that returned by uninterrupted pruning: the checkpoint mechanism changes only \emph{when} the computation executes, not \emph{what} it executes. Together with the per-vector independence established above, \sysname ensures that the final graph is returned correctly. 

\subsection{Overrun-Bounded Time Budgeting}
\label{subsec:idle-time-estimation}

While decomposition fits update work into idle windows, scheduling each subtask safely requires bounding how long it may run. Since idle window lengths vary across hops, \sysname must determine a time budget for each idle window to bound the duration an update subtask may execute. Instead of predicting the exact length of the next idle window, our goal is to derive a budget $\tau_\text{est}$ such that the expected overrun (i.e., the amount by which the budget exceeds the actual idle time) remains a controlled fraction of the mean idle time. This ensures that any time taken from the search computation phase stays within a controlled tolerance.

\begin{figure}[!t]
\centering
\includegraphics[width=0.99\columnwidth]{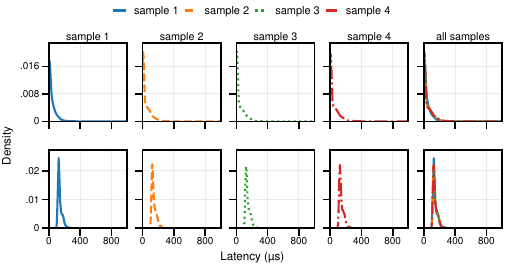}
\vspace{-6pt}
\caption{Temporal stability of idle-time distributions on SIFT-10M, including the KDEs of four independently sampled time windows and the aggregate of all samples. Top row: PipeSearch idle time; bottom row: BeamSearch I/O latency at \#IO=4.}
\label{fig:stable-distribution}
\end{figure}

Deriving such a budget requires knowledge of the idle time distribution, which is difficult to model in practice: whether it is disk I/O latency in BeamSearch or CPU idle time in PipeSearch, empirical distributions often exhibit irregular, heavily skewed shapes. Despite this, our observations reveal a useful property: within short, continuous time windows, the shape of the idle time distribution remains remarkably stable. As shown in Figure~\ref{fig:stable-distribution}, the distributional shape within each configuration is highly consistent across independently sampled time chunks. This temporal stability means that a budget derived from recent samples reliably bounds overrun for upcoming windows, enabling a {\em data-driven} approach that solves for a budget directly from the empirical distribution. 


We call a time budget \emph{overrun-bounded} if its expected overrun, as a fraction of the mean idle time, stays within a user-specified threshold. Formally, let $\tau$ denote the idle time random variable with mean $\mu_{\tau}$. Over a short historical window, we collect empirical measurements $\{\tau_i\}_{i=1}^N$, which approximate the current idle time distribution by the temporal stability property. The overall latency degradation is bounded by a user-specified threshold $\theta$ (\S\ref{subsec:design-overview}), which also bounds the I/O-window overrun. Our objective is to find a time budget $\tau_{\text{est}}$ such that the expected overrun does not exceed a fraction $\theta$ of the mean idle time:
\[
\mathbb{E}[\max(0, \tau_{\text{est}} - \tau)] \le \mu_{\tau} \cdot \theta.
\]

\sysname bounds the expected overrun by $\mu_{\tau} \cdot \theta$ to ensure that any overrun into the search phase remains a controlled fraction of the idle budget. Replacing the expectation with empirical sample means yields a solvable constraint:
\[
\frac{1}{N} \sum\nolimits_{i=1}^{N} \max(0, \tau_{\text{est}} - \tau_i) \le \frac{1}{N} \sum\nolimits_{i=1}^{N} \tau_i \cdot \theta.
\]

Since the left-hand side is monotonically non-decreasing in $\tau_{\text{est}}$, \sysname solves the empirical constraint efficiently via binary search over $\tau_{\text{est}}$ using recent idle time samples. The formulation above assumes a single idle time distribution, but BeamSearch and PipeSearch produce structurally different idle time patterns, requiring the following adaptations:
\begin{itemize}[leftmargin=*,topsep=0.2em,itemsep=0.2em,parsep=0pt,partopsep=0pt]
\item
{\em BeamSearch adaptation:} BeamSearch dynamically decides how many I/O requests to issue per hop; more concurrent requests lead to longer I/O waits and hence different idle window distributions. \sysname maintains $B_u$ buckets (one per batch size $n_{\text{requests}} \in \{1, \dots, B_u\}$), each with its own sample history and solved $\tau_{\text{est}}$, so that the budget is conditioned on the actual I/O concurrency of each hop.
\item 
{\em PipeSearch adaptation:}
PipeSearch's pipeline I/O yields heavily right-skewed idle distributions (most intervals in $[0, 10]\,\mu$s), so scheduling on every interval produces an overly conservative $\tau_{\text{est}}$. \sysname adopts a \emph{$K$-sparse} strategy: update tasks run only every $K$th interval. When solving for $\tau_{\text{est}}$, overrun is counted only on scheduled intervals while the budget is averaged over all $N$ intervals, amortizing the cost and letting $\tau_{\text{est}}$ target the longer windows that actually carry work.
\end{itemize}

\subsection{Adaptive Utilization Tuning}
\label{subsec:utuner}

The per-hop budget $\tau_{\text{est}}$ controls overrun within each idle window, but cannot by itself keep end-to-end search latency degradation within the user-specified bound $\theta$: co-execution introduces additional sources of degradation (cache pollution from displaced search data and varying sensitivity across workload phases) that per-hop analysis cannot enumerate or model. Since such additional sources cannot be eliminated directly, \sysname adds an automated, end-to-end {\em feedback loop} based on the observed actual query latency, and adaptively adjusts the fraction of $\tau_{\text{est}}$ devoted to updates to keep the ratio of end-to-end latency degradation within $\theta$. It further employs {\em cache prefetching} to mitigate cache pollution at its origin, giving the loop more headroom to push $\alpha$ higher. We elaborate on both mechanisms as follows.

\para{Feedback-loop.}
\sysname continuously adjusts a {\em utilization ratio} $\alpha$ (i.e., the fraction of $\tau_{\text{est}}$ actually allocated for updates), as workload and system conditions change. Its objective is to maximize update throughput while keeping the degradation in end-to-end search latency within the user-specified bound $\theta$. It is a feedback-driven state machine taht uses periodically observed query latency (mean by default, but replaceable with P95 or P99 by substituting the observed statistic):
\begin{itemize}[leftmargin=*]
\item
{\em Step~1: Recording.} \sysname first suspends update scheduling. It then profiles search latency and idle time to establish a no-update baseline and to compute the first-time budget~$\tau_{\text{est}}$.
\item
{\em Step~2: Binary search.} Using the baseline and $\tau_{\text{est}}$ obtained in Step~1, \sysname searches for an initial feasible $\alpha$ such that $\tfrac{\texttt{query\_latency}}{\texttt{baseline}}$ is not higher than $1+\theta$. If no feasible $\alpha$ exists, \sysname disables co-execution.
\item
{\em Step~3: Steady.} Once a feasible $\alpha$ is found, \sysname enters a continuous adaptation loop. It increases $\alpha$ by a small fixed increment when the measured latency remains below $(1+\theta)$ times of the baseline to gain additional update throughput; it decreases $\alpha$ when a violation is observed (i.e., $\tfrac{\texttt{query\_latency}}{\texttt{baseline}} > 1+\theta$). The no-update baseline is periodically refreshed by temporarily suspending update scheduling for a short profiling interval.
\item
{\em Step~4: Rebaseline.} Triggered by excessive baseline drift (i.e., three consecutive violations observed in Step~3) or Search failure, this phase returns to Step~1 to re-establish the baseline and re-find a feasible $\alpha$.
\end{itemize}

\para{Cache prefetching.}
While the feedback loop can absorb cache pollution by reducing $\alpha$, doing so sacrifices update throughput. \sysname therefore prefetches (i) the query vector, (ii) the PQ distance lookup table, and (iii) visited-set metadata. Prefetches are issued both after each update execution and after I/O completion, ensuring search data is warm before distance computation resumes. By shrinking cache-induced degradation at its origin, prefetching allows \sysname to sustain a higher $\alpha$ without violating $\theta$.

\subsection{Parameter Configuration}
\label{subsec:parameter-config}

\sysname exposes only one user-facing parameter: the latency degradation bound $\theta$, which specifies the maximum search-latency overhead allowed during search--update co-execution. A smaller $\theta$ prioritizes search latency, while a larger $\theta$ allows more aggressive use of search-side idle time for updates. Given $\theta$, \sysname automatically estimates per-hop budgets and tunes utilization at runtime. The standard ANNS parameters of the underlying graph index, such as graph degree $R$, candidate pool size $L$, beam width $W$, and result size $K$, remain application-level choices for recall, latency, and throughput; \sysname does not change their semantics.

%% file: 5-implementation.tex
\section{Implementation}
\label{sec:imple}

We choose two open-source disk-based ANNS systems as backends, FreshDiskANN \cite{freshdiskann-code} and OdinANN \cite{pipeann-code}, and implement \sysname atop them. We add approximately 1,900 lines of C++ to each.  Our co-execution model includes the search-side and update-side integration.

\para{Search-side integration.} To execute update subtasks during I/O stalls, the search thread must retain CPU control rather than blocking on I/O completion. \sysname decomposes each hop's I/O into a \emph{submit--execute--poll} protocol:
\begin{itemize}[leftmargin=*]
\item
{\em Submit:} The search thread batches all frontier read requests and submits them asynchronously via \texttt{io\_uring} \cite{joshi24}. Requests that hit the user-space page cache are resolved immediately, bypassing the kernel I/O path.
\item
{\em Execute:} While I/O is in flight, the search thread consults \sysname's time budgeting logic and, if permitted, executes one update subtask from the queue, converting a CPU stall into useful computation.
\item
{\em Poll:} The thread harvests completions with a non-blocking batch poll; if requests remain outstanding, it falls back to a blocking wait.
\end{itemize}

\para{Update-side integration.} On the update side, \sysname bridges to the search side via a shared FIFO task queue that decouples task production from consumption.
\begin{itemize}[leftmargin=*]
\item
{\em Task generation:} The update thread enqueues tasks at per-vector granularity (\S\ref{subsec:update-decomposition}). Tasks employ \emph{self-re-enqueueing}: after processing one vector, the search thread that executed the task re-enqueues it if vectors remain, eliminating the need for the update thread to manage per-vector enqueuing and reducing synchronization overhead.
\item
{\em Time-budgeted execution:}
During each idle window, a search thread dequeues one task and executes it within a time budget of $\alpha \cdot \tau_{\text{est}}$, where $\alpha$ is \sysname's utilization ratio (\S\ref{subsec:utuner}). If the budget expires mid-pruning, the checkpoint mechanism (\S\ref{subsec:update-decomposition}) saves progress and re-enqueues the continuation.
\end{itemize}

%% file: 6-evaluation.tex
\section{Evaluation}
\label{sec:eval}

\subsection{Methodology}
\label{subsec:setup}

\begin{table*}[t]
  \centering
  \setlength{\tabcolsep}{3pt}
  \renewcommand{\arraystretch}{1.08}
  \caption{Exp\,\#1 (100M end-to-end): 100M-scale datasets at default settings (\S\ref{subsec:setup}).}
  \label{tab:e2e-100m}
  \vspace{-9pt}
  \small
  \begin{tabular}{l l @{\hspace{10pt}} r r @{\hspace{10pt}} >{\centering\arraybackslash}p{4.6em} >{\centering\arraybackslash}p{4.6em} >{\centering\arraybackslash}p{4.6em} @{\hspace{10pt}} >{\centering\arraybackslash}p{4.6em} >{\centering\arraybackslash}p{4.6em} >{\centering\arraybackslash}p{4.6em}}
  \toprule
  Dataset & Workload & \multicolumn{2}{c}{Speedup} & \multicolumn{3}{c}{Delete-phase latency change (\%)} & \multicolumn{3}{c}{Insert-phase latency change (\%)} \\
  \cmidrule(lr){3-4}\cmidrule(lr){5-7}\cmidrule(lr){8-10}
  & & Delete & Insert & Avg & P95 & P99 & Avg & P95 & P99 \\
  \midrule
  \multirow{4}{*}{SIFT-100M}
  & Beam+Fresh & 1.34$\times$ & 1.27$\times$ & $+$2.6$\pm$0.06 & $-$0.3$\pm$0.09 & $+$0.8$\pm$0.63 & $+$0.8$\pm$0.17 & $-$2.0$\pm$0.23 & $+$1.8$\pm$2.50 \\
  & Pipe+Fresh & 1.25$\times$ & 1.23$\times$ & $+$0.7$\pm$0.07 & $-$1.4$\pm$0.12 & $+$0.5$\pm$0.40 & $+$0.7$\pm$0.17 & $-$1.4$\pm$0.26 & $+$3.3$\pm$1.50 \\
  & Beam+Odin  & 1.20$\times$ & 1.36$\times$ & $+$7.0$\pm$0.10 & $+$3.1$\pm$0.13 & $+$5.6$\pm$1.36 & $+$6.6$\pm$0.10 & $+$8.2$\pm$0.15 & $+$8.3$\pm$0.25 \\
  & Pipe+Odin  & 1.38$\times$ & 1.51$\times$ & $-$0.6$\pm$0.12 & $-$3.5$\pm$0.18 & $-$2.3$\pm$0.75 & $+$6.8$\pm$0.16 & $+$9.5$\pm$0.34 & $+$7.7$\pm$0.74 \\
  \midrule
  \multirow{4}{*}{DEEP-100M}
  & Beam+Fresh & 1.25$\times$ & 1.18$\times$ & $+$3.4$\pm$0.06 & $-$0.7$\pm$0.09 & $-$1.2$\pm$0.59 & $+$1.6$\pm$0.17 & $-$3.0$\pm$0.20 & $+$2.3$\pm$2.25 \\
  & Pipe+Fresh & 1.23$\times$ & 1.19$\times$ & $+$0.8$\pm$0.07 & $-$2.1$\pm$0.11 & $-$3.8$\pm$0.39 & $+$1.1$\pm$0.15 & $-$1.7$\pm$0.22 & $-$0.9$\pm$1.62 \\
  & Beam+Odin  & 1.13$\times$ & 1.47$\times$ & $+$7.1$\pm$0.09 & $+$2.0$\pm$0.11 & $+$5.8$\pm$1.63 & $+$8.6$\pm$0.09 & $+$10.1$\pm$0.13 & $+$9.7$\pm$0.16 \\
  & Pipe+Odin  & 1.42$\times$ & 1.72$\times$ & $+$1.0$\pm$0.13 & $+$0.6$\pm$0.31 & $+$5.6$\pm$1.02 & $+$7.5$\pm$0.15 & $+$6.6$\pm$0.24 & $-$3.4$\pm$0.73 \\
  \midrule
  \multirow{4}{*}{SPACEV-100M}
  & Beam+Fresh & 1.48$\times$ & 1.43$\times$ & $+$0.3$\pm$0.12 & $-$1.1$\pm$0.17 & $-$2.2$\pm$1.08 & $+$0.5$\pm$0.38 & $-$0.0$\pm$0.40 & $+$8.8$\pm$5.67 \\
  & Pipe+Fresh & 1.48$\times$ & 1.46$\times$ & $+$0.1$\pm$0.10 & $-$2.3$\pm$0.12 & $-$2.5$\pm$0.50 & $+$0.4$\pm$0.31 & $-$2.2$\pm$0.35 & $+$5.1$\pm$3.35 \\
  & Beam+Odin  & 1.26$\times$ & 1.19$\times$ & $+$7.8$\pm$0.22 & $+$5.1$\pm$0.28 & $+$8.9$\pm$0.59 & $+$6.0$\pm$0.27 & $+$7.6$\pm$0.46 & $+$9.3$\pm$0.79 \\
  & Pipe+Odin  & 1.49$\times$ & 1.48$\times$ & $-$0.8$\pm$0.23 & $-$2.1$\pm$0.33 & $-$0.3$\pm$0.94 & $+$6.4$\pm$0.29 & $+$4.6$\pm$0.63 & $-$3.2$\pm$1.43 \\
  \bottomrule
  \end{tabular}
  \vspace{-6pt}
\end{table*}

\begin{table*}[t]
  \centering
  \setlength{\tabcolsep}{3pt}
  \renewcommand{\arraystretch}{1.08}
  \caption{Exp\,\#1 (10M end-to-end): 10M-scale datasets at default settings (\S\ref{subsec:setup}).}
  \label{tab:e2e-10m}
  \vspace{-9pt}
  \small
  \begin{tabular}{l l @{\hspace{10pt}} r r @{\hspace{10pt}} >{\centering\arraybackslash}p{4.6em} >{\centering\arraybackslash}p{4.6em} >{\centering\arraybackslash}p{4.6em} @{\hspace{10pt}} >{\centering\arraybackslash}p{4.6em} >{\centering\arraybackslash}p{4.6em} >{\centering\arraybackslash}p{4.6em}}
  \toprule
  Dataset & Workload & \multicolumn{2}{c}{Speedup} & \multicolumn{3}{c}{Delete-phase latency change (\%)} & \multicolumn{3}{c}{Insert-phase latency change (\%)} \\
  \cmidrule(lr){3-4}\cmidrule(lr){5-7}\cmidrule(lr){8-10}
  & & Delete & Insert & Avg & P95 & P99 & Avg & P95 & P99 \\
  \midrule
  \multirow{4}{*}{SIFT-10M}
  & Beam+Fresh & 1.65$\times$ & 1.77$\times$ & $+$2.4$\pm$0.05 & $-$0.2$\pm$0.07 & $-$1.2$\pm$0.45 & $+$2.7$\pm$0.12 & $-$1.3$\pm$0.11 & $-$1.2$\pm$1.71 \\
  & Pipe+Fresh & 1.58$\times$ & 1.67$\times$ & $+$3.7$\pm$0.06 & $+$4.2$\pm$0.14 & $+$6.7$\pm$0.46 & $+$2.8$\pm$0.15 & $+$2.0$\pm$0.34 & $+$1.7$\pm$1.55 \\
  & Beam+Odin  & 2.02$\times$ & 1.34$\times$ & $+$8.0$\pm$0.17 & $+$7.2$\pm$0.18 & $+$6.1$\pm$0.63 & $+$2.6$\pm$0.20 & $+$3.2$\pm$0.45 & $+$2.7$\pm$0.64 \\
  & Pipe+Odin  & 2.15$\times$ & 1.72$\times$ & $+$3.5$\pm$0.10 & $+$3.5$\pm$0.13 & $+$1.1$\pm$0.38 & $+$7.8$\pm$0.27 & $+$7.5$\pm$0.61 & $+$2.6$\pm$0.76 \\
  \midrule
  \multirow{4}{*}{DEEP-10M}
  & Beam+Fresh & 1.39$\times$ & 1.48$\times$ & $+$1.9$\pm$0.08 & $-$0.2$\pm$0.13 & $+$0.2$\pm$0.42 & $+$4.3$\pm$0.16 & $+$0.2$\pm$0.22 & $+$2.5$\pm$1.77 \\
  & Pipe+Fresh & 1.60$\times$ & 1.57$\times$ & $+$3.6$\pm$0.07 & $+$5.0$\pm$0.13 & $+$7.8$\pm$0.40 & $-$0.1$\pm$0.14 & $-$0.1$\pm$0.27 & $-$0.5$\pm$1.35 \\
  & Beam+Odin  & 1.70$\times$ & 1.36$\times$ & $+$6.6$\pm$0.15 & $+$4.1$\pm$0.22 & $+$5.0$\pm$0.79 & $+$4.8$\pm$0.23 & $+$4.6$\pm$0.42 & $+$4.6$\pm$0.61 \\
  & Pipe+Odin  & 2.18$\times$ & 1.84$\times$ & $+$3.0$\pm$0.12 & $+$3.4$\pm$0.15 & $+$5.2$\pm$0.62 & $+$10.2$\pm$0.31 & $+$9.2$\pm$0.57 & $+$5.0$\pm$0.54 \\
  \midrule
  \multirow{4}{*}{SPACEV-10M}
  & Beam+Fresh & 1.81$\times$ & 2.68$\times$ & $+$2.0$\pm$0.10 & $-$0.7$\pm$0.11 & $-$0.4$\pm$0.51 & $+$7.3$\pm$0.67 & $+$2.6$\pm$1.12 & $+$8.2$\pm$4.53 \\
  & Pipe+Fresh & 1.83$\times$ & 1.90$\times$ & $+$2.0$\pm$0.11 & $+$2.4$\pm$0.43 & $+$2.1$\pm$0.57 & $+$1.4$\pm$0.29 & $+$0.8$\pm$0.86 & $+$1.7$\pm$2.25 \\
  & Beam+Odin  & 1.86$\times$ & 1.21$\times$ & $+$9.6$\pm$0.71 & $+$6.0$\pm$0.65 & $+$6.3$\pm$1.45 & $+$1.8$\pm$0.43 & $+$2.1$\pm$1.20 & $+$4.5$\pm$1.97 \\
  & Pipe+Odin  & 2.16$\times$ & 1.63$\times$ & $+$4.6$\pm$0.26 & $+$5.3$\pm$0.35 & $+$7.0$\pm$0.61 & $+$5.5$\pm$0.47 & $+$6.5$\pm$1.21 & $+$4.5$\pm$1.38 \\
  \bottomrule
  \end{tabular}
  \vspace{-6pt}
\end{table*}

\para{Testbeds.} We use two server testbeds with complementary hardware profiles. \emph{Testbed~A} is equipped with 2$\times$ 20-core Intel Xeon Gold 5218R @ 2.10\,GHz, 128\,GiB DRAM (8$\times$16\,GiB DDR4-2666), an Intel P4610 2.9\,TiB SSD, and Ubuntu 20.04.4 LTS (Linux 5.4.0-80-generic). Its 40 physical cores provide a high degree of parallelism needed to saturate both search and update threads at large scale, in which we allocate 64 search threads and 16 update threads. \emph{Testbed~B} is equipped with 2$\times$ 8-core Intel Xeon Silver 4309Y @ 2.80\,GHz, 256\,GiB DRAM (16$\times$16\,GiB DDR4-2666), a WD Ultrastar DC SN640 7\,TiB SSD, and Ubuntu 22.04.5 LTS (Linux 5.15.0-163-generic). Its smaller core count is representative of a resource-constrained deployment and is better suited for isolating per-mechanism effects, in which we allocate 28 search threads and 4 update threads. Hyper-threading is enabled on both testbeds. 

\para{Workloads.}
We evaluate mixed search-update co-execution on three datasets: SIFT \cite{jegou11}, SPACEV \cite{microsoft20}, and DEEP \cite{babenko16}, sliced at scales from 10M to 500M vectors. Unless otherwise stated, each run deletes 5\% of vectors from the original index and then inserts 5\% new vectors, with concurrent search.

\para{Default settings.} We use backend-specific latency thresholds unless otherwise stated: $\theta{=}5\%$ for FreshDiskANN-based backends and $\theta{=}8\%$ for OdinANN-based backends. The higher threshold for OdinANN reflects its user-space cache, which amplifies per-episode overhead and requires a larger budget for meaningful speedup. All indexes are built with $R=96$, $L=100$, and $B=32$. 
In all experiments, both search and update threads are continuously saturated, ensuring that no thread-level idleness exists; the only idle resource exploited by \sysname is the \emph{intra-request} I/O stall time within each busy search thread. All search operations use $K=10$ and $L=100$, and the resulting recall@10 ranges from 95\% to 99\% across all datasets. We report search-side average/P95/P99 latency and throughput (queries per second (QPS)), and update-side throughput (vectors/second), along with execution times for deletion and insertion. We collect all search-run samples, averaging 1,633 samples per run at 10M scale and 5,157 samples per run at 100M scale, and report 95\% confidence intervals using the normal distribution.

\para{Baselines.} We derive the baselines from DiskANN \cite{jayaram19} (BeamSearch), FreshDiskANN \cite{singh21} (Buffered Insert), PipeANN \cite{guo25} (PipeSearch), and OdinANN \cite{guo26} (Direct Insert), forming four configurations: (i) BeamSearch + Buffered Insert, (ii) BeamSearch + Direct Insert, (iii) PipeSearch + Buffered Insert, and (iv) PipeSearch + Direct Insert. For each configuration, we compare with \sysname against without \sysname.

\subsection{End-to-End Performance}
\label{subsec:e2e}

\textbf{Exp\,\#1 (End-to-end performance).} We evaluate all configurations at the default settings (\S\ref{subsec:setup}) on 100M-scale and 10M-scale datasets. Table~\ref{tab:e2e-100m} reports results on SIFT-100M, DEEP-100M, and SPACEV-100M using Testbed~A; Table~\ref{tab:e2e-10m} reports SIFT-10M, DEEP-10M, and SPACEV-10M using Testbed~B.

\sysname consistently reduces update execution time across all datasets and configurations. On the 100M-scale datasets, \sysname achieves an update speedup of up to 1.48$\times$ (delete) and 1.46$\times$ (insert) over the non-\sysname baseline under FreshDiskANN, and up to 1.49$\times$ (delete) and 1.72$\times$ (insert) under OdinANN. On the 10M-scale datasets, the gains are even larger: \sysname achieves an update speedup of up to 1.83$\times$ (delete) and 2.68$\times$ (insert) over the non-\sysname baseline under FreshDiskANN, and up to 2.18$\times$ (delete) and 1.84$\times$ (insert) under OdinANN. The two search strategies generally show opposite speedup rankings across the two backends, particularly during the delete phase. Under FreshDiskANN, BeamSearch tends to achieve higher speedup because its blocking I/O creates longer idle windows for \sysname to exploit. Under OdinANN, the ranking reverses: by default, OdinANN's PipeSearch does not cache disk pages fetched on cache misses; the resulting extra misses raise PipeSearch's idle ratio above BeamSearch's, yielding more co-execution opportunities and thus higher speedup.

Search latency remains controlled. On the 100M-scale datasets, mean latency overhead stays within 8.6\%, and on the 10M-scale datasets within 10.2\%; P95 remains within 10.1\% and 9.2\%, respectively. Note that a few mean-latency entries are above the configured $\theta$ due to \sysname's best-effort design: $\theta$ guides adaptive tuning but is not a hard bound for every search sample. Short workload or I/O fluctuations can, in a few cases, increase the phase average beyond the target, while most configurations remain within the budget as \sysname reduces co-execution aggressiveness. P95 and P99 sometimes fall \emph{below} the baseline, which we attribute to cache prefetching after co-execution episodes. In a few configurations, tail latency increases more noticeably (e.g., $+$9.7\% P99 for DEEP-100M Beam/OdinANN insert) because \sysname reports mean latency by default; deployments prioritizing tail latency can report P95 or P99 in the observed latency statistics with comparable throughput (Exp\,\#9).

\subsection{Mechanism Validation}
\label{subsec:mechanism-validation}

We validate the two control mechanisms that enable the default operating point: overrun-bounded time budgeting, which bounds the per-hop overrun, and adaptive utilization tuning, which adapts the usable fraction of the estimated idle budget to enforce end-to-end latency targets.

\begin{figure}[t]
  \centering
  \setlength{\tabcolsep}{2pt}
  \renewcommand{\arraystretch}{1.0}
  \begin{tabular}{@{\ }c@{\ }c@{\ }c@{\ }}
  \multicolumn{3}{c}{\includegraphics[width=1.6in]{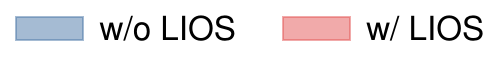}} \\
  \includegraphics[height=0.773in]{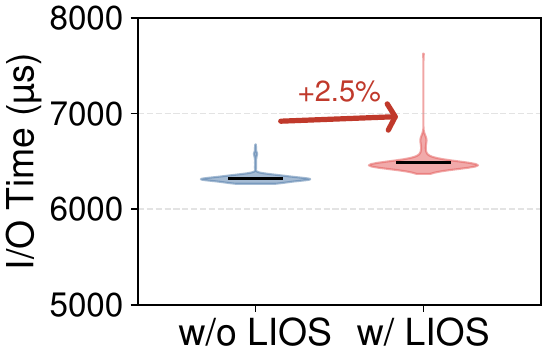} &
  \includegraphics[height=0.773in]{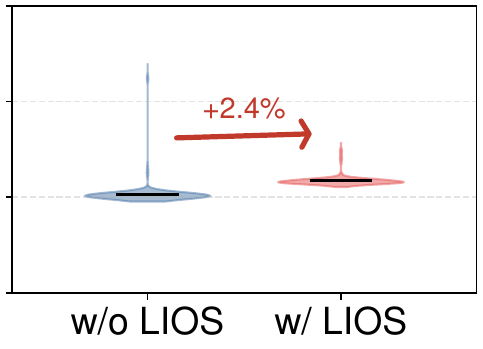} &
  \includegraphics[height=0.773in]{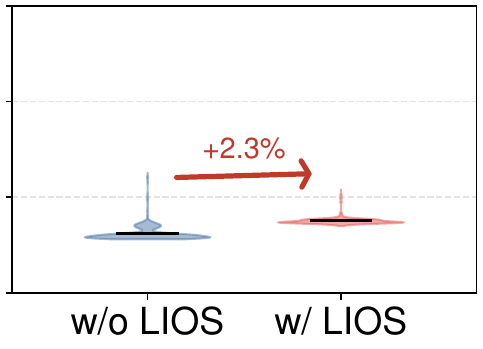} \\
  \parbox[c]{1.186in}{\centering\small (a) Beam, $R{=}32$} &
  \parbox[c]{1.072in}{\centering\small (b) Beam, $R{=}64$} &
  \parbox[c]{1.072in}{\centering\small (c) Beam, $R{=}96$} \\[4pt]
  \includegraphics[height=0.773in]{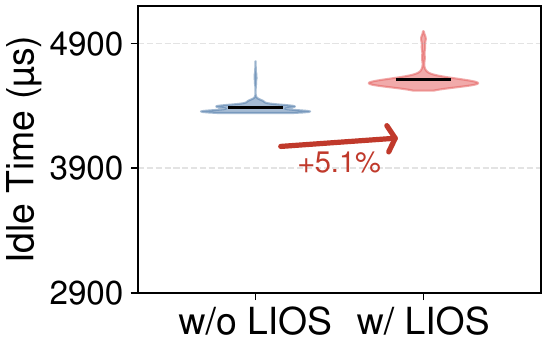} &
  \includegraphics[height=0.773in]{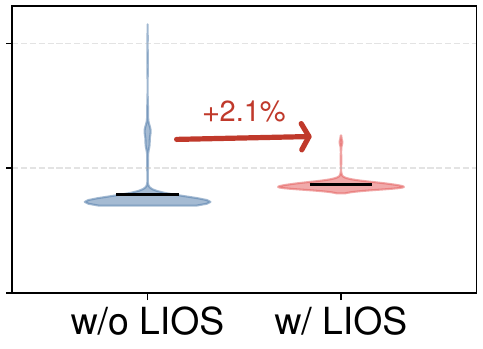} &
  \includegraphics[height=0.773in]{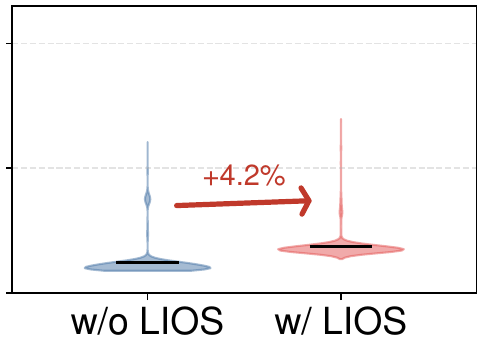} \\
  \parbox[c]{1.186in}{\centering\small (d) Pipe, $R{=}32$} &
  \parbox[c]{1.072in}{\centering\small (e) Pipe, $R{=}64$} &
  \parbox[c]{1.072in}{\centering\small (f) Pipe, $R{=}96$} \\
  \end{tabular}
  \vspace{-9pt}
  \caption{Exp\,\#2 (Overrun-bounded time budgeting): SIFT-10M, Testbed~B; $\theta{=}5\%$. Top row: BeamSearch (I/O time); bottom row: PipeSearch (idle time).}
  \label{fig:estimator-overrun-violin-delete-merge-perR}
  \vspace{-6pt}
\end{figure}

\para{Exp\,\#2 (Overrun-bounded time budgeting).} Figure~\ref{fig:estimator-overrun-violin-delete-merge-perR} shows hop-level violin distributions of per-request I/O time (BeamSearch) and idle time (PipeSearch) with and without \sysname at $R \in \{32,64,96\}$, using SIFT-10M on Testbed~B (28 search + 4 update threads, $\theta{=}5\%$). Each violin shows the mean and distribution; its width indicates sample density at each I/O or idle-time value. Under BeamSearch, \sysname increases mean per-request I/O time by 2.3--2.5\% across different $R$; under PipeSearch, the corresponding mean idle-time increases are 2.1--5.1\%. In all cases, the overhead stays at or near the $\theta{=}5\%$ bound, confirming that the overrun-bounded sampling (\S\ref{subsec:idle-time-estimation}) effectively limits per-hop overrun.

\begin{figure*}[t]
  \centering
  \includegraphics[width=\textwidth]{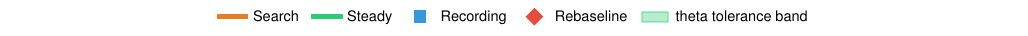}\\[-6pt]
  \includegraphics[width=\textwidth]{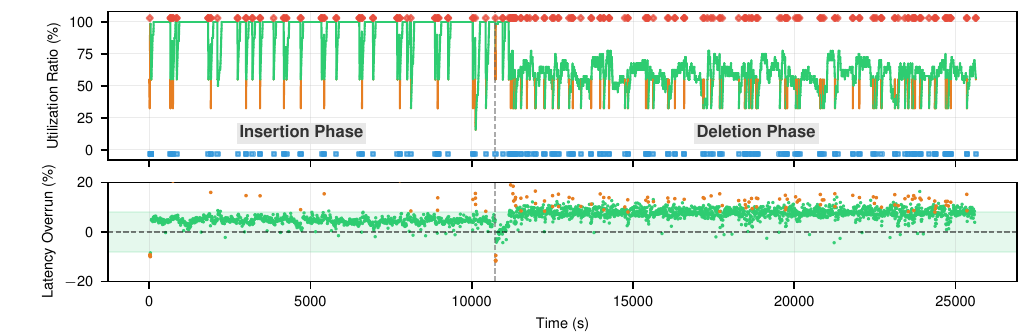}\\[-4pt]
  \vspace{-9pt}
  \caption{Exp\,\#3 (Adaptive utilization tuning): Utilization ratio and latency overrun over time on SIFT-100M.}
  \label{fig:utuner-traces}
  \vspace{-6pt}
\end{figure*}

\para{Exp\,\#3 (Adaptive utilization tuning).} We trace \sysname's utilization tuning on SIFT-100M (Testbed~A). Figure~\ref{fig:utuner-traces} shows insertion followed by deletion. \sysname reaches Steady 20s after updates start and re-enters Steady 26s after deletion begins. During insertion, it uses 94.0\% average utilization with 4.5\% latency overrun and 596 adjustments. During deletion, latency is more sensitive, so \sysname lowers utilization to 59.5\%, keeps latency overrun at 7.5\% within the $\theta{=}8\%$ target, and makes 2,127 adjustments. The few deletion samples above $\theta$ reflect transient deviations from the profiled baseline; \sysname therefore treats $\theta$ as a best-effort latency target rather than a per-sample hard bound. When such violations appear, \sysname reduces utilization and, if needed, re-enters profiling/binary search to keep phase-level latency close to the target.

\subsection{Breakdown Analysis}
\label{subsec:breakdown}

\textbf{Exp\,\#4 (Breakdown analysis).} We isolate the contribution of each major component by starting from naive co-execution and adding the mechanisms cumulatively:

\begin{itemize}[leftmargin=*,nosep]
  \item \textbf{w/o intra-vector decomposition}: only inter-vector decomposition; updates cannot be interrupted mid-pruning.
  \item \textbf{w/o overrun-bounded time budgeting}: replace \sysname's adaptive time budgeting with a fixed time budget.
  \item \textbf{w/o utilization tuner}: use a fixed utilization ratio instead of \sysname's adaptive control.
  \item \textbf{w/o cache prefetching}: disable prefetching of search-critical data after each co-execution episode.
\end{itemize}

\begin{figure}[t]
  \centering
  \setlength{\tabcolsep}{2pt}
  \renewcommand{\arraystretch}{1.0}
  \begin{tabular}{@{}c@{}}
  \includegraphics[width=\linewidth]{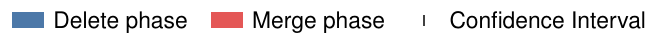} \\[2pt]
  \end{tabular}\\[-2pt]
  \begin{tabular}{@{\ }c@{\ }c@{\ }}
  \includegraphics[width=1.65in]{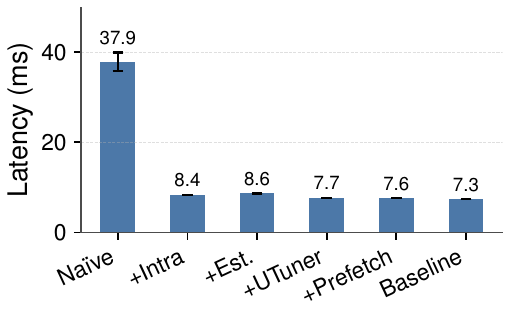} &
  \includegraphics[width=1.65in]{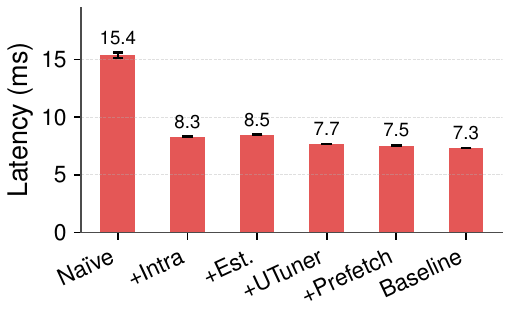} 
  \vspace{-3pt}\\
  \parbox[c]{1.58in}{\centering\small (a) Delete-phase latency} &
  \parbox[c]{1.58in}{\centering\small (b) Insert-phase latency} \\[4pt]
  \includegraphics[width=1.65in]{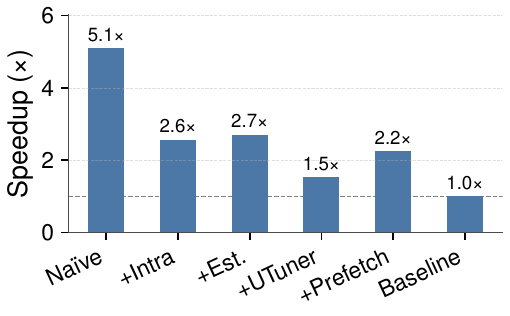} &
  \includegraphics[width=1.65in]{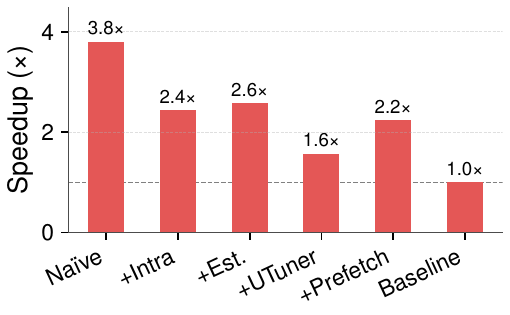} 
  \vspace{-3pt}\\
  \parbox[c]{1.58in}{\centering\small (c) Delete speedup} &
  \parbox[c]{1.58in}{\centering\small (d) Insert speedup}
  \end{tabular}
  \vspace{-9pt}
  \caption{Exp\,\#4 (Breakdown analysis): SIFT-10M ($R{=}32$, $\theta{=}5\%$). Components added cumulatively from naive co-execution. (a,\,b)~Search latency; (c,\,d)~update speedup vs.\ w/o \sysname.}
  \label{fig:breakdown}
  \vspace{-6pt}
\end{figure}

Figure~\ref{fig:breakdown} shows the results. Starting from naive co-execution without any control, the system achieves 5.1$\times$ delete speedup and 3.8$\times$ insert speedup, but at the cost of severely degraded search performance: average latency increases to 5.2$\times$ baseline during delete and 2.1$\times$ baseline during insert.
Adding \emph{intra-vector decomposition} (+Intra) enables fine-grained preemption, cutting delete-phase latency by 78\% and insert-phase latency by 46\% while retaining 2.6$\times$ and 2.4$\times$ update speedup.
Replacing the fixed time budget with \emph{overrun-bounded time budgeting} (+Est.) marginally improves speedup to 2.7$\times$/2.6$\times$ with similar latency.
Introducing \emph{adaptive utilization tuning} (+Tuning) to adaptively enforce the latency threshold brings average latency within 5\% of baseline, but reduces speedup to 1.5$\times$/1.6$\times$ as the controller conservatively limits co-execution.
Finally, enabling \emph{cache prefetching} (+Prefetch) recovers speedup to 2.2$\times$ for both phases while keeping latency within 3\% of baseline. Together, the four mechanisms achieve substantial update speedup while keeping search latency tightly controlled.

\subsection{Sensitivity and Generalization}
\label{subsec:sensitivity}

We study how the throughput-latency trade-off varies with key operating parameters, including the latency threshold~$\theta$, thread configuration, graph degree~$R$, tuning latency statistics, and dataset scale.

\begin{figure*}[t]
  \centering
  \setlength{\tabcolsep}{2pt}
  \renewcommand{\arraystretch}{1.0}
  \begin{tabular}{@{\ }c@{\ }c@{\ }c@{\ }}
  \multicolumn{3}{c}{\includegraphics[width=3.5in]{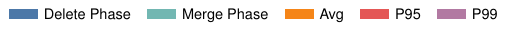}} \\

  \includegraphics[width=2.20in]{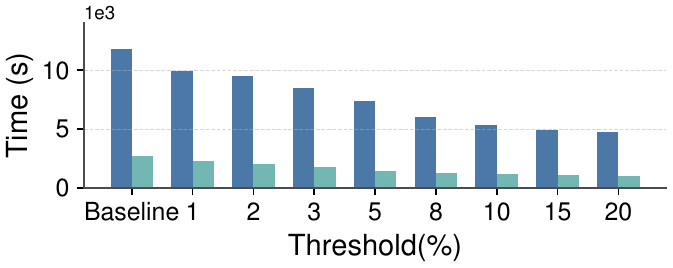} &
  \includegraphics[width=2.20in]{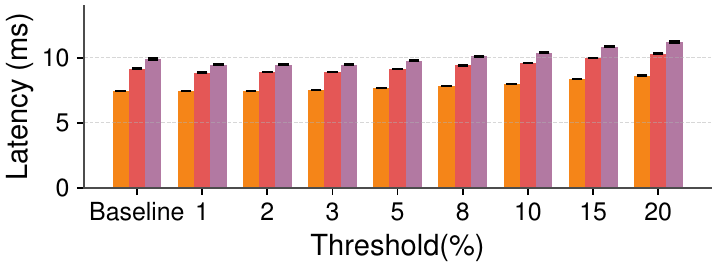} &
  \includegraphics[width=2.20in]{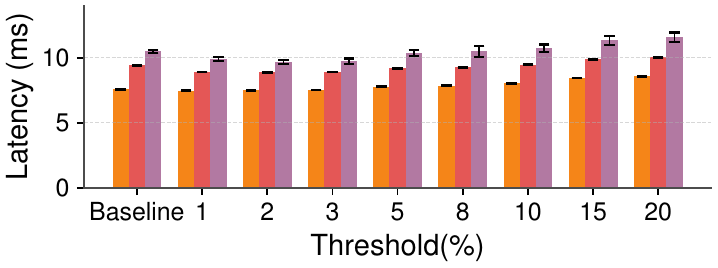} \\
  \parbox[c]{2.10in}{\centering\small (a) Beam: Execution time} &
  \parbox[c]{2.10in}{\centering\small (b) Beam: Delete-phase latency} &
  \parbox[c]{2.10in}{\centering\small (c) Beam: Insert-phase latency} \\

  \includegraphics[width=2.20in]{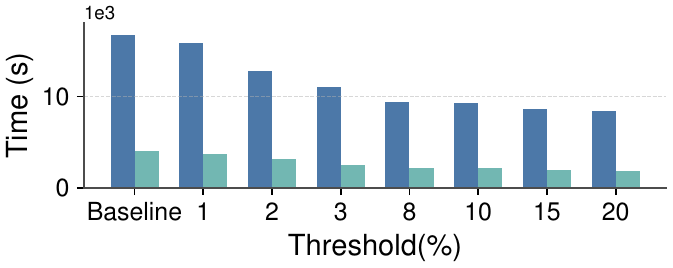} &
  \includegraphics[width=2.20in]{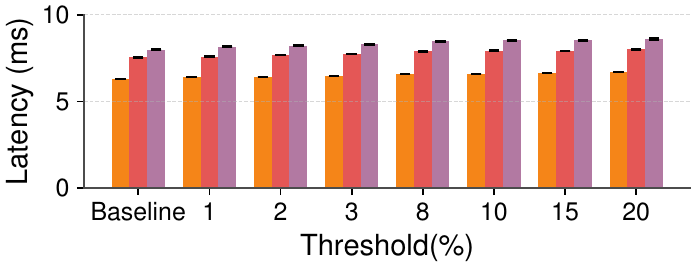} &
  \includegraphics[width=2.20in]{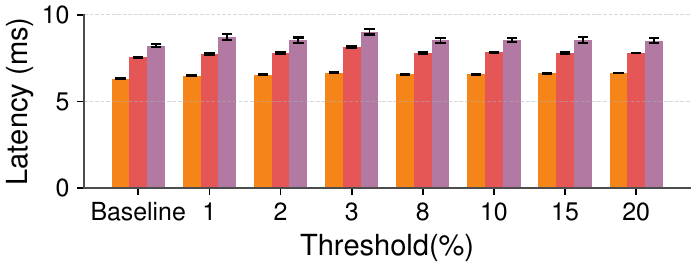} \\
  \parbox[c]{2.10in}{\centering\small (d) Pipe: Execution time} &
  \parbox[c]{2.10in}{\centering\small (e) Pipe: Delete-phase latency} &
  \parbox[c]{2.10in}{\centering\small (f) Pipe: Insert-phase latency} \\
  \end{tabular}
  \vspace{-9pt}
  \caption{Exp\,\#5 (Latency threshold sensitivity): SIFT-10M, Testbed~B; $R{=}96$. Top row: BeamSearch; bottom row: PipeSearch.}
  \label{fig:sensitivity-threshold}
  \vspace{-6pt}
\end{figure*}

\begin{figure*}[t]
  \centering
  \setlength{\tabcolsep}{2pt}
  \renewcommand{\arraystretch}{1.0}
  \begin{tabular}{@{\ }c@{\ }c@{\ }c@{\ }}
  \multicolumn{3}{c}{\includegraphics[width=3.5in]{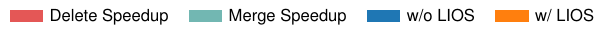}} \\

  \includegraphics[width=2.20in]{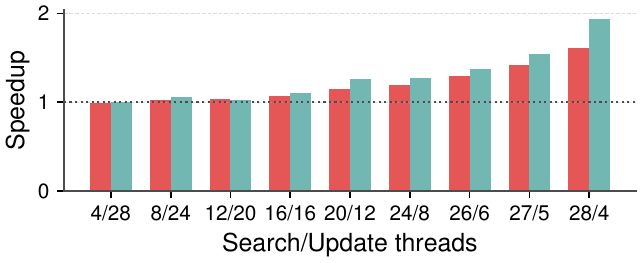} &
  \includegraphics[width=2.20in]{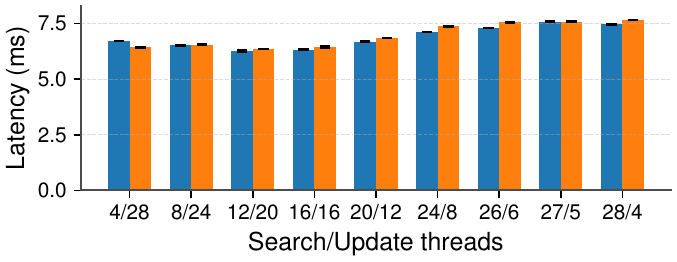} &
  \includegraphics[width=2.20in]{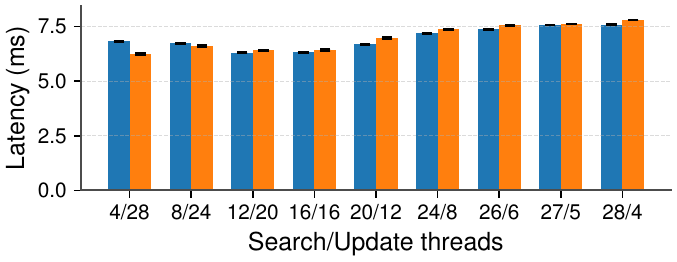} \\
  \parbox[c]{2.10in}{\centering\small (a) Beam: Speedup} &
  \parbox[c]{2.10in}{\centering\small (b) Beam: Delete-phase latency} &
  \parbox[c]{2.10in}{\centering\small (c) Beam: Insert-phase latency} \\

  \includegraphics[width=2.20in]{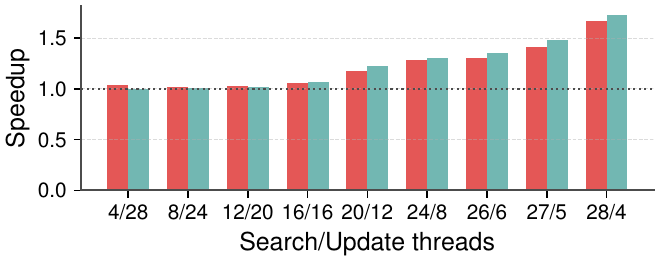} &
  \includegraphics[width=2.20in]{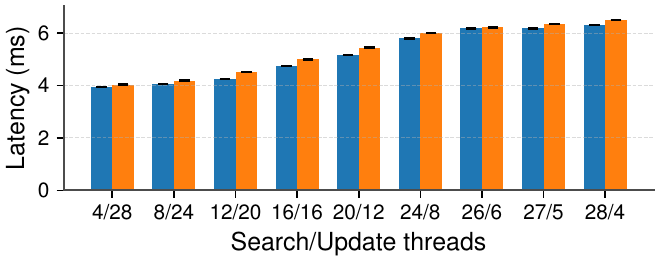} &
  \includegraphics[width=2.20in]{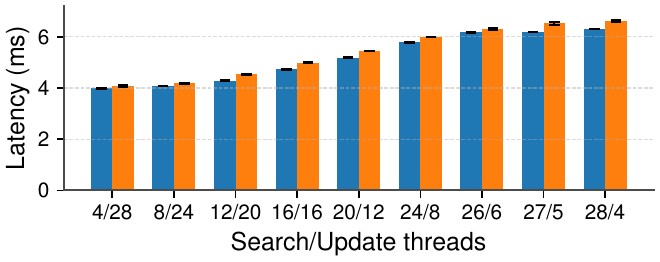} \\
  \parbox[c]{2.10in}{\centering\small (d) Pipe: Speedup} &
  \parbox[c]{2.10in}{\centering\small (e) Pipe: Delete-phase latency} &
  \parbox[c]{2.10in}{\centering\small (f) Pipe: Insert-phase latency} \\
  \end{tabular}
  \vspace{-9pt}
  \caption{Exp\,\#6 (Thread configuration sensitivity): SIFT-10M, Testbed~B; $R{=}96$, $\theta=5\%$. Top: BeamSearch; bottom: PipeSearch. Left: speedup; right: mean search latency (w/o \sysname vs.\ w/ \sysname).}
  \label{fig:sensitivity-thread}
  \vspace{-6pt}
\end{figure*}

\para{Exp\,\#5 (Latency threshold sensitivity).} Figure~\ref{fig:sensitivity-threshold} characterizes the configurable throughput-latency frontier for BeamSearch and PipeSearch. Under BeamSearch, \sysname achieves an update-time reduction of up to 53\% (delete) and 54\% (insert) over the non-\sysname baseline when sweeping $\theta$ from 1\% to 20\%. At the default $\theta=5\%$, delete time already decreases by 38\% and insert time by 48\%, while mean search latency grows by less than 3\%.

Under PipeSearch, \sysname achieves an update-time reduction of up to 47\% (delete) and 49\% (insert) over the non-\sysname baseline under the same sweep. Even at conservative thresholds, co-execution produces meaningful time savings; across the full 1\%--20\% sweep, mean latency growth stays under 7\%.

\para{Exp\,\#6 (Thread configuration sensitivity).} Different deployments allocate varying fractions of CPU resources to search vs.\ update depending on workload priorities: a search-heavy service uses more search threads, while an ingestion-heavy pipeline favors update threads. Figure~\ref{fig:sensitivity-thread} evaluates \sysname across such configurations. Gains grow as a larger share of threads is allocated to search: at the default 28/4 search/update split, \sysname achieves an update-time reduction of up to 38\% (delete) and 48\% (insert) over the non-\sysname baseline under BeamSearch, up from negligible at the 4/28 split. PipeSearch exhibits the same pattern, reaching 40\% and 42\% at the 28/4 split. This is expected: more search threads generate more I/O idle periods, creating additional overlap opportunities for \sysname. Notably, \sysname provides meaningful speedup across all tested thread ratios, confirming its applicability to a wide range of workload mixes.

\para{Exp\,\#7 (Graph degree sensitivity).} Table~\ref{tab:sensitivity-graph-degree} shows that co-execution gains diminish as graph degree increases. For BeamSearch, \sysname achieves an update-time reduction of up to 55\% (delete) and 57\% (insert) over the non-\sysname baseline at $R{=}32$, which narrows to 38\% and 48\% at $R{=}96$. PipeSearch follows the same trend: the reduction decreases from 56\% to 40\% for delete and from 55\% to 42\% for insert. The insert-side decline is milder because insert operations are intrinsically less compute-intensive than delete processing. Across all configurations, QPS drop remains under 4.1\% and search latency growth stays within the target budget.

\begin{table*}[t]
  \centering
  \setlength{\tabcolsep}{5pt}
  \renewcommand{\arraystretch}{1.08}
  \caption{Exp\,\#7 (Graph degree sensitivity): SIFT-10M, Testbed~B; 28/4 threads, $\theta=5\%$.}
  \label{tab:sensitivity-graph-degree}
  \vspace{-9pt}
  \small
  \begin{tabular*}{\textwidth}{@{\extracolsep{\fill}}lccccccccc@{}}
  \toprule
  Workload & $R$ & \multicolumn{2}{c}{Speedup} & \multicolumn{3}{c}{Delete phase} & \multicolumn{3}{c}{Insert phase} \\
  \cmidrule(lr){3-4}\cmidrule(lr){5-7}\cmidrule(lr){8-10}
  &  & Delete & Insert & \makecell[c]{Time (s)\\w/o$\rightarrow$w/} & \makecell[c]{QPS$\downarrow$ (\%)} & \makecell[c]{Lat$\uparrow$ (ms)} & \makecell[c]{Time (s)\\w/o$\rightarrow$w/} & \makecell[c]{QPS$\downarrow$ (\%)} & \makecell[c]{Lat$\uparrow$ (ms)} \\
  \midrule
  Beam & 32 & 2.23$\times$ & 2.34$\times$ & 1987.7$\rightarrow$892.9 & 2.59$\pm$0.11 & +0.202$\pm$0.009 & 568.3$\rightarrow$242.8 & 2.67$\pm$0.48 & +0.207$\pm$0.042 \\
  Beam & 64 & 1.82$\times$ & 2.12$\times$ & 7456.9$\rightarrow$4096.8 & 2.92$\pm$0.25 & +0.252$\pm$0.028 & 1790.4$\rightarrow$845.7 & 3.46$\pm$0.15 & +0.290$\pm$0.013 \\
  Beam & 96 & 1.61$\times$ & 1.94$\times$ & 11793.8$\rightarrow$7349.2 & 2.61$\pm$0.08 & +0.209$\pm$0.009 & 2735.3$\rightarrow$1412.9 & 2.54$\pm$0.19 & +0.217$\pm$0.018 \\
  \midrule
  Pipe & 32 & 2.29$\times$ & 2.24$\times$ & 2739.7$\rightarrow$1199.0 & 4.08$\pm$0.08 & +0.318$\pm$0.007 & 783.3$\rightarrow$349.6 & 3.71$\pm$0.23 & +0.291$\pm$0.018 \\
  Pipe & 64 & 1.93$\times$ & 2.08$\times$ & 10603.8$\rightarrow$5507.1 & 2.08$\pm$0.11 & +0.149$\pm$0.011 & 2663.9$\rightarrow$1279.8 & -0.06$\pm$0.27 & -0.010$\pm$0.021 \\
  Pipe & 96 & 1.67$\times$ & 1.73$\times$ & 16666.3$\rightarrow$10006.1 & 2.92$\pm$0.07 & +0.213$\pm$0.006 & 3984.0$\rightarrow$2303.2 & 3.78$\pm$0.43 & +0.317$\pm$0.045 \\
  \bottomrule
  \end{tabular*}
  \vspace{-6pt}
\end{table*}

This trend is expected: larger $R$ increases per-update computation, including more distance evaluations and graph maintenance, while search-side I/O idle time does not grow with $R$. Each update must therefore be split into more slices to fit within the same idle windows, raising scheduling overhead and reducing the net benefit.

\begin{table*}[t]
  \centering
  \small
  \setlength{\tabcolsep}{3pt}
  \renewcommand{\arraystretch}{1.08}
  \caption{Exp\,\#8 (Dataset scale): SIFT at 10M--500M, OdinANN backend, default settings (\S\ref{subsec:setup}).}
  \label{tab:dataset-scale-sift-odin}
  \vspace{-9pt}
  \begin{tabular}{l l @{\hspace{10pt}} r r @{\hspace{10pt}} >{\centering\arraybackslash}p{4.6em} >{\centering\arraybackslash}p{4.6em} >{\centering\arraybackslash}p{4.6em} @{\hspace{10pt}} >{\centering\arraybackslash}p{4.6em} >{\centering\arraybackslash}p{4.6em} >{\centering\arraybackslash}p{4.6em}}
  \toprule
  Dataset & Workload & \multicolumn{2}{c}{Speedup} & \multicolumn{3}{c}{Delete-phase latency change (\%)} & \multicolumn{3}{c}{Insert-phase latency change (\%)} \\
  \cmidrule(lr){3-4}\cmidrule(lr){5-7}\cmidrule(lr){8-10}
  & & Delete & Insert & Avg & P95 & P99 & Avg & P95 & P99 \\
  \midrule
  SIFT-10M & Beam (OdinANN) & 1.19$\times$ & 1.21$\times$ & $+$6.9$\pm$0.36 & $+$3.6$\pm$0.53 & $+$8.0$\pm$4.76 & $+$4.8$\pm$0.29 & $+$6.1$\pm$0.38 & $+$9.2$\pm$4.84 \\
  SIFT-20M & Beam (OdinANN) & 1.17$\times$ & 1.26$\times$ & $+$7.0$\pm$0.21 & $+$3.6$\pm$0.40 & $+$7.2$\pm$3.07 & $+$4.4$\pm$0.23 & $+$5.6$\pm$0.32 & $+$8.7$\pm$2.47 \\
  SIFT-50M & Beam (OdinANN) & 1.18$\times$ & 1.28$\times$ & $+$6.6$\pm$0.14 & $+$3.2$\pm$0.19 & $+$6.8$\pm$1.93 & $+$5.9$\pm$0.15 & $+$7.5$\pm$0.21 & $+$7.5$\pm$0.34 \\
  SIFT-100M & Beam (OdinANN) & 1.20$\times$ & 1.36$\times$ & $+$7.0$\pm$0.10 & $+$3.1$\pm$0.13 & $+$5.6$\pm$1.36 & $+$6.6$\pm$0.10 & $+$8.2$\pm$0.15 & $+$8.3$\pm$0.25 \\
  SIFT-200M & Beam (OdinANN) & 1.27$\times$ & 1.13$\times$ & $+$6.7$\pm$0.08 & $+$2.9$\pm$0.10 & $+$7.0$\pm$1.08 & $+$3.0$\pm$0.12 & $+$3.2$\pm$0.17 & $+$4.6$\pm$0.29 \\
  SIFT-500M & Beam (OdinANN) & 1.16$\times$ & 1.04$\times$ & $+$6.7$\pm$0.07 & $+$2.6$\pm$0.07 & $+$4.9$\pm$0.90 & $+$1.8$\pm$0.07 & $+$2.0$\pm$0.09 & $+$2.4$\pm$0.14 \\
  \bottomrule
  \end{tabular}
  \vspace{-6pt}
\end{table*}

\para{Exp\,\#8 (Dataset scale sensitivity).}
We sweep SIFT from 10M to 500M vectors on Testbed~A. As shown in Table~\ref{tab:dataset-scale-sift-odin}, delete-phase speedup remains consistent across all scales (1.16\,--\,1.27$\times$). Insert-phase speedup is lower at 200M (1.13$\times$) and 500M (1.04$\times$) than at 10M\,--\,100M (1.21\,--\,1.36$\times$); this is an artifact of sharded index construction, where memory constraints force multi-shard builds that leave many nodes with degree well below~$R$, replacing full \textsc{RobustPrune} with inexpensive appends and thus reducing the CPU-intensive work available for \sysname; profiling confirms the pruning ratio drops from 100\% at 100M to 3.6\% and 12.4\% at 200M and 500M. Mean latency overhead stays within 7\% across all scales, confirming that \sysname's effectiveness is scale-independent.

\vspace{-1.0em}
\subsection{Additional Experiments}
\label{subsec:additional-experiments}

We include two additional studies in the Appendix in the supplementary file. Exp\,\#9 evaluates the latency statistics used by the utilization tuner and shows that adding P95 or P99 to the observed signal tightens latency control while preserving comparable delete speedup. Exp\,\#10 examines resource utilization and shows that \sysname does not increase search-side disk traffic, roughly doubles effective CPU utilization, and adds only KiB-level checkpoint memory overhead.

%% file: 7-relatedwork.tex
\section{Related Work}
\label{sec:related-work}

DiskANN \cite{jayaram19} and SPANN \cite{chen21} are two pioneering disk-based ANNS systems. DiskANN adopts a graph-based index inspired by prior in-memory graph-based indexes (e.g., NSG \cite{fu19} and HNSW \cite{malkov18}), while SPANN adopts a cluster-based index that organizes vectors into partitions for disk-efficient retrieval. \sysname specifically targets graph-based ANNS systems (including DiskANN and its variants) as their updates are compute-bound and can effectively fill CPU idle time during search I/O stalls. SPANN's updates, by contrast, are I/O-bound, leaving limited room for such co-execution. 

Several follow-up studies of DiskANN focus on optimizing \emph{search performance}, primarily by reducing disk I/Os or hiding I/O latency. One line of research improves data layout and locality to minimize redundant reads during graph traversal: Starling \cite{wang24} co-optimizes data layout and search flows to shorten traversal paths; PageANN \cite{kang25} introduces a page-aligned graph layout to ensure that each hop maps to a single SSD page read; Gorgeous \cite{yin25} revisits caching and layout designs to achieve higher throughput with fewer disk accesses. VeloANN \cite{zhao26} combines locality-aware layout, buffer pooling, and coroutine-based async execution to reduce storage stalls. Another line of research aims to entirely mask I/O latency: PipeANN \cite{guo25} pipelines computation and SSD I/Os to narrow the gap between disk-based and in-memory searches; Li et al. \cite{li26} survey memory, disk, and algorithm-layer designs targeting this I/O bottleneck. However, these search-centric optimizations primarily focus on accelerating query execution, often leaving residual CPU idle time during unresolved I/O bottlenecks unexploited.

Some ANNS systems target \emph{update performance}. FreshDiskANN \cite{singh21} buffers inserts and deletes in a short-term in-memory index and periodically merges them into a long-term SSD index at a cost proportional only to the change set. SPFresh \cite{xu23} builds on SPANN with a cluster-partitioned index and incremental vector reassignment, achieving update throughput comparable to full rebuilds with lower overhead than graph-based methods. IP-DiskANN \cite{xu25} supports direct in-place updates on the graph, avoiding periodic merges while preserving recall under streaming updates. OdinANN \cite{guo26} preserves stable search performance under dynamic vector insertions through an in-place insertion strategy.

\sysname focuses on \emph{resource utilization} rather than optimizing search or updates in isolation. It schedules update work to run during I/O stalls in search threads, exploiting otherwise wasted CPU idle time. This co-execution strategy improves resource utilization without requiring new index structures or sacrificing search latency.

%% file: 8-conclusion.tex
\section{Conclusion}
\label{sec:conclusion}

\sysname is a search-update co-execution framework for improving resource efficiency in disk-based graph ANNS systems. By harnessing intra-request CPU idle time during search-side I/O stalls, \sysname runs fine-grained update work without reducing the CPU budget for search. Its resumable update decomposition, overrun-bounded time budgeting, and adaptive utilization tuning fit update tasks into short, variable stall windows while controlling search latency. Evaluations show that \sysname accelerates updates by up to 2.68$\times$ in insertion and 2.18$\times$ in deletion while keeping search latency bounded.

%% file: appendix.tex
\section{Additional Experiments}
\label{app:additional-experiments}

We report two additional experiments on SIFT-10M using Testbed~B. We first study how the observed latency statistics affect the online tuner's behavior, then measure disk I/O, CPU utilization, and checkpoint memory overhead.

\para{Exp\,\#9 (Observed latency statistics).} We compare three choices of observed latency statistics (mean only, mean+P95, and mean+P95+P99) under the default BeamSearch configuration. As shown in Table~\ref{tab:utuner-signal-tail}, incorporating P95 notably tightens mean-latency control (overhead drops from $+$2.975\% to $+$1.879\%) while further reducing tail latency. Adding P99 yields diminishing returns because P99 samples are sparser and noisier, making online adjustment less stable. Delete speedup stays comparable across all three settings (1.88--1.94$\times$), confirming that tail-aware observed statistics curb burst-driven overscheduling without sacrificing overlap opportunity.

\begin{table}[H]
  \centering
  \small
  \setlength{\tabcolsep}{4pt}
  \renewcommand{\arraystretch}{1.08}
  \caption{Exp\,\#9 (Observed latency statistics): SIFT-10M, BeamSearch + FreshDiskANN, Testbed~B; 28/4 threads, $\theta{=}5\%$. Results averaged over $R \in \{32,64,96\}$.}
  \label{tab:utuner-signal-tail}
  \vspace{-9pt}
  \begin{tabular}{lcccc}
  \toprule
  \makecell[l]{Observed latency\\statistics} & Delete spd. & Mean lat. & P95 lat. & P99 lat. \\
  \midrule
  Mean & 1.88$\times$ & $+$2.975\% & $-$1.747\% & $-$2.481\% \\
  Mean+P95 & 1.94$\times$ & $+$1.879\% & $-$2.744\% & $-$3.144\% \\
  Mean+P95+P99 & 1.92$\times$ & $+$1.747\% & $-$2.911\% & $-$3.738\% \\
  \bottomrule
  \end{tabular}
  \vspace{-6pt}
\end{table}

\begin{figure}[H]
  \centering
  \begin{tabular}{@{}c@{}}
  \includegraphics[width=1.75in]{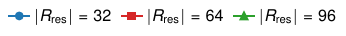} \\[-1pt]
  \includegraphics[width=2.55in]{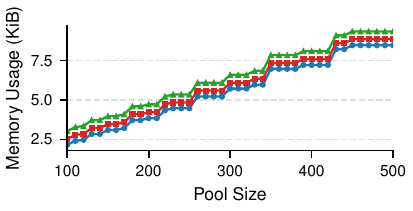} \\
  \end{tabular}
  \vspace{-9pt}
  \caption{Exp\,\#10 (Resource utilization): Checkpoint memory overhead on SIFT-10M/Testbed~B.}
  \label{fig:memory-overhead}
  \vspace{-6pt}
\end{figure}

\para{Exp\,\#10 (Resource utilization analysis).} We measure three resources to understand where \sysname's gains come from: search-side disk I/O, CPU utilization, and checkpoint memory. Table~\ref{tab:disk-io-avg} first shows that \sysname does not increase search-side disk traffic: BeamSearch has identical average I/O counts with and without \sysname (119.17 I/Os/query), while PipeSearch changes only from 120.38 to 120.20 I/Os/query ($-$0.15\%). Table~\ref{tab:cpu-util-avg} then shows that \sysname converts idle/polling windows into productive update work, roughly doubling effective utilization across both phases (93--128\% improvement). Finally, Figure~\ref{fig:memory-overhead} shows that this added state is small: each checkpoint remains at the KiB level; for example, selecting $R{=}96$ from a candidate pool of size 500 requires less than 10\,KiB.

\begin{table}[H]
  \centering
  \small
  \setlength{\tabcolsep}{4pt}
  \renewcommand{\arraystretch}{1.08}
  \caption{Exp\,\#10 (Resource utilization): Average disk I/Os per query on SIFT-10M/Testbed~B.}
  \label{tab:disk-io-avg}
  \vspace{-9pt}
  \begin{tabular}{lccc}
  \toprule
  Search Algo. & \makecell{Avg. I/Os (w/o)} & \makecell{Avg. I/Os (w/)} & Change (\%) \\
  \midrule
  BeamSearch & 119.17 & 119.17 & +0.00 \\
  PipeSearch & 120.38 & 120.20 & -0.15 \\
  \bottomrule
  \end{tabular}
  \vspace{-6pt}
\end{table}

\begin{table}[H]
  \centering
  \small
  \setlength{\tabcolsep}{3pt}
  \renewcommand{\arraystretch}{1.08}
  \caption{Exp\,\#10 (Resource utilization): Phase-wise CPU utilization on SIFT-10M/Testbed~B ($R{=}96$, 28/4 threads, $\theta{=}5\%$). w/o = without \sysname; w/ = with \sysname.}
  \label{tab:cpu-util-avg}
  \vspace{-9pt}
  \begin{tabular}{lcccc}
  \toprule
  Search Algo. & Phase & Util. (w/o) & Util. (w/) & \makecell{Improvement\\(\%)} \\
  \midrule
  BeamSearch & Delete & 0.269 & 0.535 & +98.6 \\
  BeamSearch & Insert & 0.264 & 0.603 & +128.3 \\
  PipeSearch & Delete & 0.362 & 0.728 & +101.4 \\
  PipeSearch & Insert & 0.368 & 0.713 & +93.7 \\
  \bottomrule
  \end{tabular}
  \vspace{-6pt}
\end{table}

